\documentclass[preprint,floats,tightenlines,11pt,eqsecnum,showpacs,aps]{revtex4}
\usepackage{amsmath}
\usepackage{graphicx}
\begin{document}
\def\appls{\hbox{$<$\kern-.75em\lower 1.00ex\hbox{$\sim$}}}

\title{EVIDENCE FOR $\rho^0(770)-f_0(980)$ MIXING IN $\pi^- p \to \pi^-\pi^+n$ \\ FROM CERN MEASUREMENTS ON POLARIZED TARGET}

\author{Miloslav Svec\footnote{electronic address: svec@hep.physics.mcgill.ca}}
\affiliation{Physics Department, Dawson College, Montreal, Quebec, Canada H3Z 1A4}
\date{November 17, 2014}

\begin{abstract}

We present model independent high resolution amplitude analyses of CERN data on $\pi^- p \to \pi^- \pi^+ n$ at 17.2 GeV/c for dipion masses 580-1080 MeV at small momentum transfers on polarized target (Analysis I) and unpolarized target (Analysis II). The results for $S$- and $P$-wave transversity amplitudes from the Analysis I are used in a model independent determination of helicity amplitudes. All three analyses provide similar evidence for $\rho^0(770)-f_0(980)$ mixing in $S$-and $P$-wave amplitudes. In Analyses I and II the $S$-wave amplitudes $|S_d|^2$ peak at $\rho^0(770)$ mass while the $P$-wave amplitudes $|L_d|^2$ dip at $f_0(980)$ mass. In both analyses the $S$-wave amplitudes $S_\tau$ are nearly in phase with longitudinal amplitudes $L_\tau$ and nearly $180^\circ$ out of phase with transverse amplitudes $U_\tau$ indicating resonant phases of $\rho^0(770)$ also in the $S$-wave for both target nucleon transversities $\tau=u,d$. The $S$-wave single flip helicity amplitudes $|S_1|^2$ and the relative phases $\Phi(L_1)-\Phi(S_1)$ show the same pattern. While the non-flip helicity amplitudes $|L_0|^2$ are small below 960 MeV, there is a sudden rise above this mass that may reflect the presence of $f_0(980)$ resonance. We show that the apparent dependence of $\rho^0(770)$ width on helicity in amplitudes $|U_\tau|^2$ and $|N_\tau|^2$ results from the interference of transversity amplitudes with definite dipion helicities $\lambda =\pm1$ and not from the violation of rotational/Lorentz symmetry. The consistency of $\rho^0(770)-f_0(980)$ mixing with this symmetry suggests that the production mechanism and the mechanism responsible for $\rho^0(770)-f_0(980)$ mixing have two independent dynamical origins. 

\end{abstract}

\maketitle

\tableofcontents

\newpage
\section{Introduction.}
Following the discovery in 1961 of $\rho$ meson in $\pi N \to \pi \pi N$ reactions~\cite{erwin61}, the measurements of forward-backward asymmetry in $\pi^- p \to \pi^- \pi^+n$ suggested the existence of a rho-like resonance in the $S$-wave, later referred to as $\sigma(750)$ scalar meson~\cite{hagopian63,islam64,patil64,durand65,baton65}. It was expected that  $\sigma(750)$ would show up prominently in $\pi^- p \to \pi^0 \pi^0 n$ production where $\rho^0(770)$ does not contribute. The measurements of this reaction at CERN in 1972 found no evidence for a rho-like $\sigma(750)$~\cite{apel72}. Furthermore, in 1973, Pennington and Protopopescu used analyticity and unitarity constraints on partial wave amplitudes in $\pi \pi \to \pi \pi$ scattering (Roy equations) to show that a narrow $\sigma(750)$ resonance cannot contribute to $\pi \pi$ scattering~\cite{pennington73}. From these facts it was concluded that $\sigma(750)$ does not exist in $\pi^- p \to \pi^- \pi^+ n$ and in 1974 Particle Data Group dropped this state from its listings. 

In 1978 Lutz and Rybicki showed that almost complete amplitude analysis of reactions $\pi N \to \pi^+ \pi^- N$ is possible from measurements in a single experiment on a transversely polarized target~\cite{lutz78}. CERN-Munich-Cracow group measured $\pi^- p \to \pi^-\pi^+n$ on transversely polarized proton target in a high statistics experiment at 17.2 GeV/c at low momentum transfers $-t=0.005 - 0.20$ (GeV/c)$^2$~\cite{becker79a,becker79b,chabaud83} and at high momentum transfers $-t=0.20 - 1.00$ (GeV/c)$^2$~\cite{rybicki85}. Saclay group measured $\pi^+ n \to \pi^+ \pi^- p$ on transversely polarized deuteron target at 5.98 and 11.85 GeV/c at larger momentum transfers $-t = 0.20 - 0.40$ (GeV/c)$^2$~\cite{lesquen85,svec92a}. Measurements of $\pi^- p \to \pi^-\pi^+n$ on transversely polarized proton target at 1.78 GeV/c at low momentum transfers $-t=0.005 - 0.20$ (GeV/c)$^2$ were made at ITEP~\cite{alekseev99}.

The CERN measurements of $\pi^- p \to \pi^-\pi^+n$ and $\pi^+ n \to \pi^+ \pi^- p$ on polarized targets reopened the question of the existence of $\sigma(750)$ scalar meson. Evidence for a narrow $\sigma(750)$ was found in amplitude analyses of $\pi^- p \to \pi^-\pi^+n$ at 17.2 GeV/c~\cite{donohue79,becker79b,rybicki85,svec92c} and in $\pi^+ n \to \pi^+ \pi^- p$ at 5.98 and 11.85~\cite{svec84,svec92c}. Clear evidence for $\sigma(750)$ emerged from later and more precise amplitude analyses of both reactions~\cite{svec96,svec97a,svec02a}. Another evidence for $\sigma(750)$ comes from the amplitude analysis of the ITEP data at 1.78 GeV/c~\cite{alekseev99}. 

In 2001, E852 Collaboration at BNL reported high statistics measurements $\pi^- p \to \pi^0 \pi^0 n$ at 18.3 GeV/c~\cite{gunter01}. The data revealed large differences in the $S$-wave intensities in $\pi^- \pi^+$ and $\pi^0 \pi^0$ production. There was no evidence for a rho-like $\sigma(750)$ resonance in the $\pi^0 \pi^0$ $S$-wave. The BNL data presented anew the puzzle of the $\sigma(750)$ resonance. The CERN and BNL data are both high quality data that cannot be used to exclude one another. We must accept that they are both correct and that the apparent contradictions between them are telling us something new and important. The puzzle of $\sigma(750)$ resonance has become a unique opportunity to learn new physics.

The first hints on the solution of the puzzle emerged from the amplitude analysis of CERN measurements of $\pi^- p \to \pi^- \pi^+ n$ in Ref.~\cite{svec02a}. This study extended the dipion mass range of 600 - 900 MeV of previous analyses~\cite{svec96,svec97a} to 580 - 1080 MeV to include the $f_0(980)$ resonance. The $S$-wave transversity ampltudes $S_\tau$ with target nucleon trasversity $\tau=u (up),d (down)$  were fitted using a parametrization with Breit-Wigner amplitudes for both resonances and complex backgoudnds. The best fit to all moduli of $S$-wave transversity amplitudes gives $m_\sigma=778 \pm 16$ MeV and $\Gamma_\sigma = 142 \pm 33$ MeV. These values are very close to resonance parameters of $\rho^0(770)$ with $m_\rho = 775.49 \pm 0.34$ MeV and $\Gamma_\rho = 149.1 \pm 0.8$ MeV. Both solutions for the $P$-wave amplitude $|L_d|^2$ with zero dipion helicity showed an unexpected dip at $\sim$ 980 MeV, the mass of $f_0(980)$ resonance. 

This work suggested that the dip in $|L_d|^2$ arises from $\rho^0(770)-f_0(980)$ mixing in the $P$-wave. Similar mixing in the $S$-wave then allows us to identify $\sigma(750)$ with $\rho^0(770)$ which immediately explains why no $\sigma(750)$ was ever found in $\pi^- p \to \pi^0 \pi^0 n$ production. The $S$-wave amplitudes are either in phase or 180$^\circ$ out of phase with the resonating $P$-wave amplitudes which is in agreement with the hypothesis of $\rho^0(770)-f_0(980)$ mixing.  

In Ref.~\cite{svec02a} we used the results of the fits to the transversity amplitudes to calculate $S$-wave helicity non-flip and flip amplitudes $S_0$ and $S_1$ assuming the residues of $\sigma(750)$ were real and positive in both transversity amplitudes and there was no additional relative phase apart from the fits. The results suggested that $\sigma(750)$ dominates $S_0$ while it is suppressed in $S_1$, and therefore that it dominates in $\pi \pi \to a_1 \pi$ and is suppressed in $\pi \pi \to \pi \pi$ scattering. However, the results for $S_1$ disagreed with established $\pi \pi$ scattering amplitudes. In over 100 later fits with different input values of complex residues of $\sigma(750)$ and other fitting parameters we obtained mostly convergent fits. They all had the same $\chi^2$ and values of resonance parameters of $\sigma(750)$ and $f_0(980)$ identical to the values from our original fits. However, the corresponding aplitudes $S_0$ and $S_1$ were widely different with many of them unphysical solutions. We must conclude that the presence of a rho-like resonance in $S$-wave transversity amplitudes is not related to $\pi \pi \to a_1 \pi$ or $\pi \pi \to \pi \pi$ scattering. It stands on its own and lends support to the hypothesis of $\rho^0(770)-f_0(980)$ mixing.

In this work we present our results of a high resolution amplitude analysis of CERN data on polarized target (Analysis I) and of a new model independent amplitude analysis of CERN data on unpolarized target (Analysis II). We trace the presence of $\rho^0(770)$ in the $S$-wave transversity amplitudes in both Analyses to the data component $a_1+a_2$ in the equation $|S|^2=a_1+a_2-3|L|^2$. The data $a_1+a_2$ have a large $\rho^0(770)$ peak that is not balanced out by $\rho^0(770)$ peak in the $P$-wave amplitude $|L|^2$. The relative phases between $S$-wave amplitudes and $P$-wave amplitudes $L_\tau$ and $U_\tau$ are near $0^\circ$ and $-180^\circ$, respectively in both Analyses. Since the $P$-wave amplitudes have $\rho^0(770)$ Breit-Wigner phase, the $S$-wave amplitudes must have nearly the same resonant phase.

Next we outline a model independent method to determine the $S$- and $P$-wave helicity amplitudes from the knowledge of transversity amplitudes and present new results on helicity amplitudes $S_n$ and $L_n$, $n=0,1$. The four solutions for the single flip amplitude $|S_1|^2$ show a clear $\rho^0(770)$ peak and the phase of $S_1$ is again nearly equal to the phase of $P$-wave  amplitude $L_1$. There is a remarkable similarity and concordance between the Analyses I, II and the helicity amplitudes. 

At first sight the hypothesis of $\rho^0(770)-f_0(980)$ mixing suggests an apparent violation of rotational and Lorentz symmetry in $\pi^- p \to \pi^- \pi^+ n$ which appeared to be supported by the observation of large differences in the $\rho^0(770)$ widths in the moduli of $P$-wave amplitues $|U_\tau|^2$ and $|N_\tau|^2$ with transverse dipion helicities. We show that the differences in the apparent widths of $|U_\tau|^2$ and $|N_\tau|^2$ amplitudes arise from the interference of transversity amplitudes of opposite dipion helicities $\lambda=\pm1$, and that the width of $\rho^0(770)$ resonance does not depend on its helicity in both Analyses I and II. This means that the $\rho^0(770)-f_0(980)$ mixing is fully consistent with the rotational/Lorentz symmetry of the production process.

The evidence for $\rho^0(770)-f_0(980)$ mixing does not change by including $D$-wave amplitudes in the amplitude analysis. In a related paper~\cite{svec12d} we show that the $\rho^0(770)-f_0(980)$ mixing is fully consistent with the presence of $D$-wave amplitudes assuming that $D$-wave amplitudes below 745 MeV are smaller than those above 745 MeV. In that work we also show the experimental consistency of $\rho^0(770)-f_0(980)$ mixing with isospin relations between $S$-wave amplitudes in the $\pi^-p \to \pi^- \pi^+ n$, $\pi^-p \to \pi^0 \pi^0 n$ and $\pi^+p \to \pi^+ \pi^+ n$ processes.

The presented results are in agreement with all other amplitude analyses from five different measurements of pion production on polarized target. For a full review of the evidence for $\rho^0(770)-f_0(980)$ mixing from these other analyses see Ref.~\cite{svec12d}.

The paper is organized as follows. In Section II. we define the $S$-and $P$-wave subsystem and the reduced transversity amplitudes used in all our analyses. In Sections III. and IV. we present the analytical solutions for transversity amplitudes with polarized and unpolarized target data which define the Analyses I and II, respectively. Evidence for the $\rho^0(770)-f_0(980)$ mixing from Analyses I and II is presented in Section V. In Section VI. we outline our method of the model independent determination of $S$-and $P$-wave helicity amplitudes and present evidence for  $\rho^0(770)-f_0(980)$ mixing from amplitudes $|S_1|^2$ and phases $\Phi(L_1)-\Phi(S_1)$. In Section VII. we use Analyses I and II to demonstrate the helicity independence of the $\rho^0(770)$ width and thus Lorentz symmetry of the production mechanism. The interpretation of the $\rho^0(770)-f_0(980)$ spin mixing is discussed in Section VIII. The paper closes with our conclusions and outlook in Section IX. In the Appendix we show that the measurements of $Im \rho_x^0$ and $Im \rho_z^0$ determine uniquely the signs of relative phases in the Analyses I and II.

\section{$S$- and $P$-wave subsystem of reduced final state density matrix.}

The final state density matrix in $\pi^- p \to \pi^- \pi^+ n$ has a general form~\cite{lutz78,svec12b}
\begin{eqnarray}
\rho_f(\theta \phi, \vec{P}) & = & {1 \over{2}} (I^0 (\theta \phi, \vec{P}) \sigma^0 + {\vec{I}} (\theta \phi, \vec{P}) \vec{\sigma})\\
\nonumber
& = & {1 \over{2}} \bigl (1+\vec{Q}(\theta \phi, \vec{P}) \vec{\sigma}\bigr ) I^0 (\theta \phi, \vec{P})
\end{eqnarray}
where $\theta, \phi$ describe the direction of $\pi^-$ in the dipion centre-of-mass system and $\vec{P}$ and $\vec{Q}(\theta\phi,\vec{P})$ are polarization vectors of the target and recoil nucleon, respectively. The most feasible experiments are measurements on unpolarized or polarized targets leaving the recoil nucleon polarization vector unobserved. Such measurements provide information on the reduced final state density matrix $I^0(\theta \phi,\vec{P})$ given by~\cite{lutz78,svec12b} 
\begin{equation}
I^0(\theta \phi,\vec{P}) = Tr(\rho_f(\theta \phi,\vec{P}))=
I^0_u(\theta \phi)+P_x I^0_x(\theta\phi) + P_y I^0_y(\theta\phi) + P_z I^0_z(\theta\phi)
\end{equation}
where the polarization components of $I^0(\theta \phi,\vec{P})$ are expressed in terms of density matrix elements 
\begin{equation}
I^0_k(\theta \phi) = {{d^2 \sigma} \over{dtdm}}
\sum \limits_{J \leq J'}^{J_{max}} \sum \limits _{\lambda \geq 0} \sum \limits_{\lambda'} 
\xi_{JJ'} \xi_\lambda (Re \rho^0_k)^{JJ}_{\lambda \lambda'} Re(Y^J_\lambda(\theta\phi)Y^{J*}_{\lambda'}(\theta \phi))
\end{equation}
for k=u (unpolarized target) and k=y (transversely polarized taget normal to the scattering plane), and
\begin{equation}
I^0_k(\theta \phi) = {{d^2 \sigma} \over{dtdm}} 
\sum \limits_{ J \leq J'}^{J_{max}} \sum \limits_{\lambda \geq 0} \sum \limits_{\lambda'} 
\xi_{JJ'} \xi_\lambda (Im \rho^0_k)^{JJ'}_{\lambda \lambda'} 
Im(Y^J_\lambda(\theta \phi)Y^{J'*}_{\lambda'}(\theta \phi))
\end{equation}
for k=x (transversely polarized target in the scattering plane) and k=z (longitudinally polarized target). In (2.3) and (2.4) $\xi_0=1$ and $\xi_\lambda=2$ for $\lambda >0$ and the factor $\xi_{JJ'}=1$ for $J=J'$ and $\xi_{JJ'}=2$ for $J<J'$. In a given region of dipion mass $m$ and momentum transfer $t$ only amplitudes with $J \leq J_{max}$ contribute and all sums in (2.3) and (2.4) are finite. From the measured intensity $I^0(\theta \phi, \vec{P})$ the density matrix elements are determined in small $(m,t)$ bins using maximum likelihood method~\cite{eadie83,grayer74,becker79a,lesquen85}.

Amplitude analyses of measurements on polarized targets are best performed using nucleon transversity amplitudes with definite $t$-channel naturality $U^J_{\lambda, \tau}$ and $N^J_{\lambda, \tau}$ corresponding to unnatural and natural exchange, respectively~\cite{lutz78,svec12b}. $J$ and $\lambda$ are dimeson spin and helicity and $\tau=u,d$ is the nucleon transversity for target spin "up" or "down" relative to the scattering plane. In $\pi^- p \to \pi^- \pi^+n$ the amplitudes $U^J_{\lambda,\tau}$ exchange $\pi$ and $a_1$ quantum numbers in the $t$-channel while the amplitudes $N^J_{\lambda, \tau}$ exchange $a_2$ quantum numbers. General expressions for density matrix elements in terms of transversity amplitudes were tabulated by Lutz and Rybicki~\cite{lutz78} and are reproduced in the Ref.~\cite{svec12b} in our notation. For $S$- and $P$-wave transversity amplitudes we shall use a simplified notation $A_\tau$ where the amplitudes $A=S,L,U,N$ are defined as
\begin{equation}
S_\tau=U^0_{0,\tau}, \quad L_\tau=U^1_{0,\tau}
\end{equation}
\[
U_\tau=U^1_{1,\tau}, \quad N_\tau=N^1_{1,\tau}
\]
Amplitude analysis of the $S$- and $P$-wave subsystem of the reduced density matrix determines reduced transversity amplitudes defined as follows
\begin{equation}
S=|S_u|, \quad \overline {S} = |S_d|
\end{equation}
\[
L=|L_u| \exp i \left (\Phi_{L_u}-\Phi_{S_u} \right), \quad
\overline {L}=|L_d| \exp i \left ( \Phi_{L_d}-\Phi_{S_d} \right )
\]
\[
U=|U_u| \exp i \left (\Phi_{U_u}-\Phi_{S_u} \right). \quad
\overline {U}=|U_d| \exp i \left ( \Phi_{U_d}-\Phi_{S_d} \right )
\]
\[
N=|N_u| \exp i \left (\Phi_{N_u}-\Phi_{S_d} \right), \quad
\overline {N}=|N_d| \exp i \left ( \Phi_{N_d}-\Phi_{S_u} \right )
\]
where $\Phi_{A_\tau}$ is the phase of the amplitude $A_\tau$. The reduced transversity amplitudes are related to transversity amplitudes by phase factors
\begin{equation}
A_u=A \exp {i \Phi_{S_u}}, \qquad A_d=\overline {A} \exp {i \omega} \exp {i \Phi_{S_u}}
\end{equation}
for unnatural exchange amplitudes $A=S,L,U$ and  
\begin{equation}
N_u=N \exp {i \omega} \exp {i \Phi_{S_u}}, \qquad N_d=\overline {N} \exp {i \Phi_{S_u}}
\end{equation}
for natural exchange amplitude $N$. In (2.7) and (2.8) $\Phi_{S_u}$ is the arbitrary absolute phase and $\omega=\Phi_{S_d}-\Phi_{S_u}$ is the relative phase between $S$-wave amplitudes of opposite transversity.
\begin{table}
\caption{Density matrix elements $Re \rho^0_u {{d^2 \sigma}/{dtdm}}$ and $Re \rho^0_y {{d^2 \sigma}/{dtdm}}$ in terms of reduced transversity amplitudes.}
\begin{tabular}{ccc}
\toprule
$\rho^{JJ'}_{\lambda \lambda}$&$Re \rho^0_u {{d^2 \sigma}/{dtdm}}$&$Re \rho^0_y {{d^2 \sigma} /{dtdm}}$\\
\colrule
$\rho^{00}_{ss}$&$|S|^2+|\bar {S}|^2$&$|S|^2-|\bar {S}|^2$\\
$\rho^{11}_{00}$&$|L|^2+|\bar {L}|^2$&$|L|^2-|\bar {L}|^2$\\
$\rho^{11}_{11}$&${1\over{2}}(|U|^2+|\bar {U}|^2)+{1\over{2}}(|N|^2+|\bar {N}|^2)$&
${1\over{2}}(|U|^2-|\bar {U}|^2)+{1\over{2}}(|N|^2-|\bar {N}|^2)$\\
$\rho^{11}_{1-1}$&$-{1\over{2}}(|U|^2+|\bar {U}|^2)+{1\over{2}}(|N|^2+|\bar {N}|^2)$&
$-{1\over{2}}(|U|^2-|\bar {U}|^2)+{1\over{2}}(|N|^2-|\bar {N}|^2)$\\
$Re \rho^{10}_{0s}$&$Re (L S^*+\bar {L} \bar {S}^*)$&$Re (LS^*-\bar {L} \bar {S}^*)$\\
$\sqrt{2}Re \rho^{11}_{01}$&$Re (LU^*+\bar {L} \bar {U}^*)$&$Re (LU^*-\bar {L} \bar {U}^*)$\\
$\sqrt{2}Re \rho^{10}_{1s}$&$Re (US^*+\bar {U} \bar {S}^*)$&$Re (US^*-\bar {U} \bar {S}^*)$\\
\colrule
$\rho^{JJ'}_{\lambda \lambda}$&$Im \rho^0_x {{d^2 \sigma}/{dtdm}}$&$Im \rho^0_z {{d^2 \sigma} /{dtdm}}$\\
\colrule
$\sqrt{2}Im \rho^{01}_{s1}$&$Re(-S \bar {N}^*+N \bar {S}^*)$&$Im(+S \bar {N}^*-N \bar {S}^*)$\\
$\sqrt{2}Im \rho^{11}_{01}$&$Re(-L \bar {N}^*+N \bar {L}^*)$&$Im(+L \bar {N}^*-N \bar {L}^*)$\\
$Im \rho^{11}_{-11}$&$Re(+U \bar {N}^*-N \bar {U}^*)$&$Im(-U \bar {N}^*+N \bar {S}^*)$\\
\botrule
\end{tabular}
\label{Table I.}
\end{table}
In the Table I. we present $S$- and $P$-wave density matrix elements expressed in terms of the reduced transversity amplitudes. Because of the angular properties of $Y^1_\lambda(\theta \phi)$, the elements $(\rho^0_k)^{00}_{00} \equiv (\rho^0_k)^{00}_{ss}$, $(\rho^0_k)^{11}_{00}$ and $(\rho^0_k)^{11}_{11}, k=u,y$ are not independent but appear as two independent combinations in the measured angular distributions (2.3)
\begin{equation}
(\rho^0_k)_{SP} \equiv (\rho^0_k)^{00}_{ss}+(\rho^0_k)^{11}_{00}+2(\rho^0_k)^{11}_{11}, 
\qquad
(\rho^0_k)_{PP} \equiv (\rho^0_k)^{11}_{00}-(\rho^0_k)^{11}_{11}
\end{equation}

At any dipion mass the entire $S$- and $P$-wave subsystem can be solved analytically for the transversity amplitudes $A_\tau$ up to an overall phase selected to be $\Phi_{S_u}$ - withount any measurements of recoil nucleon polarization. The observables $Re \rho^0_u$ and $Re \rho^0_y$ determine the amplitudes $S,L,U,|N|$ and $\overline{S},\overline{L}, \overline{U}, |\overline{N}|$. The observables $Im \rho^0_x$ and $Im \rho^0_z$ determine the phases of amplitudes $N$ and $\overline{N}$ (Appendix). The relative phase $\omega$ is determined analytically by the process of conversion of transversity amplitudes into helicity amplitudes (Section VII.). The amplitude analysis yields two solutions for $A_u(i),i=1,2$ and two independent solutions for $A_d(j),j=1,2$ which give rise to four different $S$-and $P$-wave final state density matrices $\rho_f(\theta \phi,\vec{P};ij)$.

\section{Amplitude analysis with polarized target data on $Re\rho^0_u$ and 
$Re\rho^0_y$.}

\subsection{The moduli of transversity amplitudes}

The measured observables $\rho^0_u$ and $\rho^0_y$ organize themselves into two groups involving amplitudes of opposite transversity. Using the expressions in Table I. and $\Sigma={{d^2 \sigma}/{dtdm}}$, the two groups are
\begin{eqnarray}
a_1 & = &{1 \over{2}}((\rho^0_u)_{SP}+(\rho^0_y)_{SP}) \Sigma=|S|^2+|L|^2+|U|^2+|N|^2\\
\nonumber
a_2 & = &((\rho^0_u)_{PP}+(\rho^0_y)_{PP}) \Sigma=2|L|^2 -|U|^2-|N|^2\\
\nonumber
a_3 &= &((\rho^0_u)^{11}_{1-1}+(\rho^0_y)^{11}_{1-1}) \Sigma=|N|^2-|U|^2\\
\nonumber
a_4 &= &{1 \over {2}}((\rho^0_u)^{10}_{0s}+(\rho^0_y)^{10}_{0s}) \Sigma=
|L||S| \cos(\Phi_{LS})\\
\nonumber
a_5 & = &{1 \over {\sqrt{2}}}((\rho^0_u)^{11}_{01}+(\rho^0_y)^{11}_{01}) \Sigma=|L||U| \cos(\Phi_{LU})\\
\nonumber
a_6 &= &{1 \over {\sqrt{2}}}((\rho^0_u)^{10}_{1s}+(\rho^0_y)^{10}_{1s}) 
\Sigma=|U||S| \cos(\Phi_{US})
\end{eqnarray}
for reduced transversity amplitudes with transversity $\tau=u$, and
\begin{eqnarray}
\overline {a}_1 & = &{1 \over{2}}((\rho^0_u)_{SP}-(\rho^0_y)_{SP}) \Sigma=
|\overline {S}|^2+|\overline {L}|^2+|\overline {U}|^2+|\overline {N}|^2\\
\nonumber
\overline {a}_2 & = &((\rho^0_u)_{PP}-(\rho^0_y)_{PP}) \Sigma=
2|\overline {L}|^2 -|\overline {U}|^2-|\overline {N}|^2\\
\nonumber
\overline {a}_3 & = &((\rho^0_u)^{11}_{1-1}-(\rho^0_y)^{11}_{1-1}) \Sigma=
|\overline {N}|^2-|\overline {U}|^2\\
\nonumber
\overline {a}_4 & = &{1 \over {2}}((\rho^0_u)^{10}_{0s}-(\rho^0_y)^{10}_{0s}) \Sigma=
|\overline {L}||\overline {S}| \cos(\overline {\Phi}_{LS})\\
\nonumber
\overline {a}_5 & = &{1 \over {\sqrt{2}}}((\rho^0_u)^{11}_{01}-(\rho^0_y)^{11}_{01}) \Sigma=
|\overline {L}||\overline {U}| \cos(\overline {\Phi}_{LU})\\
\nonumber
\overline {a}_6 & = &{1 \over {\sqrt{2}}}((\rho^0_u)^{10}_{1s}-(\rho^0_y)^{10}_{1s}) \Sigma=
|\overline {U}||\overline {S}| \cos(\overline {\Phi}_{US})
\end{eqnarray}
for reduced transversity amplitudes with transversity $\tau=d$. The relative phases are 
defined as in (2.6). For dipion masses where $S$- and $P$-wave dominate, $(\rho^0_u)_{SP}$ and $(\rho^0_y)_{SP}$ are traces
\begin{equation}
(\rho^0_u)_{SP}=Tr((\rho^0_u)^{JJ}_{\lambda \lambda})=1, \quad
(\rho^0_y)_{SP}=Tr((\rho^0_y)^{JJ}_{\lambda \lambda})=T
\end{equation}
where $T$ is target spin asymmetry.

The 6 equations (2.11) and (2.12) each involve 7 unknowns for 4 moduli and 3 cosines, and are not solvable. The missing equation in each group is supplied not by the data but by phase relations
\begin{eqnarray}
\Phi_{LS}-\Phi_{LU}-\Phi_{US} & = &
(\Phi_{L_u}-\Phi_{S_u})-(\Phi_{L_u}-\Phi_{U_u})-(\Phi_{U_u}-\Phi_{S_u})=0\\
\nonumber
\overline {\Phi}_{LS}-\overline {\Phi}_{US}-\overline {\Phi}_{LU}
 & = & (\Phi_{L_d}-\Phi_{S_d})-(\Phi_{L_d}-\Phi_{U_d})-(\Phi_{U_d}-\Phi_{S_d})=0
\end{eqnarray}
These conditions lead to non-linear relations between the cosines
\begin{eqnarray}
\cos^2(\Phi_{LS})+\cos^2(\Phi_{LU})+\cos^2(\Phi_{US})-
2\cos(\Phi_{LS})\cos(\Phi_{LU})\cos(\Phi_{US}) & = & 1\\
\nonumber
\cos^2(\overline {\Phi}_{LS})+\cos^2(\overline {\Phi}_{LU})+\cos^2(\overline {\Phi}_{US})-
2\cos(\overline {\Phi}_{LS})\cos(\overline {\Phi}_{LU})\cos(\overline {\Phi}_{US}) & = & 1
\end{eqnarray}
Substituting for the cosines from (3.1) we get
\begin{equation}
a_5^2|S|^2+a_6^2|L|^2+a_4^2|U|^2-|S|^2|L|^2|U|^2=2a_4a_5a_6
\end{equation}
From (3.1) we have three equations for moduli
\begin{eqnarray}
|S|^2 = (a_1+a_2)-3|L|^2\\
\nonumber
|U|^2=|L|^2-{1 \over {2}}(a_2+a_3)\\
\nonumber
|N|^2=|L|^2-{1 \over {2}}(a_2-a_3)
\end{eqnarray}
Substituting (3.7) into (3.6) we get a cubic equation for $|L|^2 \equiv x$
\begin{equation}
ax^3+bx^2+cx+d=0
\end{equation} 
where $a=3$ and 
\begin{eqnarray}
b & = & -3[{1 \over {3}}(a_1+a_2)+{1 \over {2}}(a_2+a_3)]\\
\nonumber
c & = &{1 \over {2}}(a_1+a_2)(a_2+a_3)+a_4^2+a_6^2-3a_5^2\\
\nonumber
d & = &(a_1+a_2)a_5^2-{1 \over {2}}(a_2+a_3)a_4^2-2a_4a_5a_6
\end{eqnarray}
From (3.2) we obtain equations for moduli $|\overline {S}|^2, |\overline {U}|^2, |\overline {N}|^2$ similar to (3.7) and from (3.5) a cubic equation for $|\overline {L}|^2 = \overline {x}$ with observables $\overline {a}_k,k=1,6$ replacing $a_k,k=1,6$ in the coefficients (3.9).

The cubic equations for $|L|^2$ and $|\overline {L}|^2$ are just another form of the phase conditions (3.4). The equations can be solved analyticaly. To solve for $x$ we write $x=y-y_0$ and require that the cubic equation (3.8) transforms to the form
\begin{equation}
y^3+3Py+2Q=0
\end{equation}
This is accomplished with $y_0=b/3a$ and
\begin{equation}
P=-y_0^2 + \Bigl ({c \over {3a}} \Bigr ), \qquad 
2Q= -y_0 \Bigl (2P+{c \over {3a}} \Bigr ) +{d \over {a}} 
\end{equation}
Next we define quantities
\begin{equation}
R=\text {sign}(Q) \sqrt{|P|}, \qquad V={Q \over {R^3}} \geq 0
\end{equation}
There are three categories of solutions of cubic equation (3.10)~\cite{svec92a,bronshtein54}. Their numerical study using a Monte Carlo method shows there are two physical solutions that have an analytical form
\begin{eqnarray}
|L(1)|^2 & = & -y_0+R \cos ({\phi \over {3}}) + \sqrt{3} R \sin({\phi \over {3}})\\
\nonumber
|L(2)|^2 & = & -y_0+R \cos ({\phi \over {3}}) - \sqrt{3} R \sin({\phi \over {3}})
\nonumber
\end{eqnarray}
where $\cos(\phi)=V$, $R>0$ and $|L(1)|^2 > |L(2)|^2 $. It follows from (3.7) that the two physical solutions for the moduli $|S(i)|^2$, $|U(i)|^2$ and $|N(i)|^2$, $i=1,2$, have a similar analytical form. Similarly, there are two independent solutions for the moduli $|\overline{A}(j)|^2$, $j=1,2$ with $R>0$ and $|\overline{L}(1)|^2 > |\overline{L}(2)|^2$.

\subsection{The relative phases of transversity amplitudes}

For each set of solutions for moduli $|A(i)|$, $i=1,2$ we calculate from (3.1) the cosines
\begin{eqnarray} 
\cos(\Phi_{LS}(i))={a_4 \over {|L(i)||S(i)|}}\\
\nonumber 
\cos(\Phi_{LU}(i))={a_5 \over {|L(i)||U(i)|}}\\
\nonumber
\cos(\Phi_{US}(i))={a_6 \over {|U(i)|}|S(i)|}\\
\nonumber
\end{eqnarray}
with similar equations for $\cos(\overline {\Phi}_{LS}(j))$, $\cos(\overline {\Phi}_{LU}(j))$, $\cos(\overline {\Phi}_{US}(j))$. From the phase condition
\begin{equation}
\Phi_L-\Phi_S=(\Phi_L-\Phi_U)+(\Phi_U-\Phi_S)
\end{equation}
we obtain 3 equations for sines
\begin{eqnarray}
C_A=\sin(\Phi_{LS}) \sin(\Phi_{LU})=+\cos(\Phi_{US})-\cos(\Phi_{LS}) \cos(\Phi_{LU})\\
\nonumber
C_B=\sin(\Phi_{LS}) \sin(\Phi_{US})=+\cos(\Phi_{LU})-\cos(\Phi_{LS}) \cos(\Phi_{US})\\
\nonumber
C_C=\sin(\Phi_{US}) \sin(\Phi_{LU})=-\cos(\Phi_{LS})+\cos(\Phi_{US}) \cos(\Phi_{LU})\\
\nonumber
\end{eqnarray}
The system is solvable provided that sign($C_C$)=sign($C_A C_B$). With 
\begin{equation}
\sin(\Phi_{LS})= \epsilon \sqrt{1-\cos^2(\Phi_{LS})}, \quad 
\end{equation}
where $\epsilon=\pm 1$, $\sin(\Phi_{LU})$ and $\sin(\Phi_{US})$ can be calculated from $C_A$ and $C_B$, respectively. This procedure allows us to determine the phases $\Phi_{LS}(i)$, $\Phi_{LU}(i)$ and $\Phi_{US}(i)$. With
\begin{equation}
\sin(\overline{\Phi}_{LS})= \overline {\epsilon} \sqrt{1-\cos^2(\overline {\Phi}_{LS})}
\end{equation}
where $\overline {\epsilon}=\pm 1$, we obtain for each set of solutions for the moduli $|\overline {A}(j)|$, $j=1,2$ the phases $\overline {\Phi}_{LS}(j)$, 
$\overline {\Phi}_{LU}(j)$, $\overline {\Phi}_{US}(j)$. The amplitude analyses assumed positive root for both sines $\sin(\Phi_{LS})$ and $\sin(\overline{\Phi}_{LS})$. Such procedure is uniqe provided that the phases $\Phi_{LS}$ and $\overline {\Phi}_{LS}$ do not change signs. A change of sign would manifest itself as a double zero in the phases $\Phi_{LS}$ and $\overline {\Phi}_{LS}$ calculated from the positive roots for the sines. We found no evidence of a double zero in the phases $\Phi_{LS}$ and $\overline {\Phi}_{LS}$ so calculated, and apart from the sign ambiguity, these phases are uniquely determined. 

The phases $\Phi_{LS}$ and $\overline {\Phi}_{LS}$ are closely related to the positivity of the reduced density matrices $\rho^0(P_y)=\rho^0_u+P_y \rho^0_y$ for pure initial states $P_y= \pm 1$ by a relation $\det(\rho^0(P_y)) \sim \sin^2(\Phi_{LS})$. The fact that these phases do not change signs implies that the eigenvalues of the density matrices $\rho^0(P_y= \pm1)$ are all non-zero.

For negative roots in (3.17) and (3.18) all phases change signs. There are four combinations of the signs $\epsilon$ and $\overline {\epsilon}$. The phases with opposite signs correspond to complex conjugate reduced transversity amplitudes. For any given solution of moduli $|A(i)|, |\overline {A}(j)|,i,j=1,2$ there is a four-fold ambiguity in the phases of the reduced transversity amplitudes $A=L,U$
\begin{equation}
L(i,\epsilon)=|L(i)|\exp (i\epsilon \Phi_{LS}(i)), \quad 
\overline {L}(j,\overline{\epsilon})=|\overline {L}(j)|\exp (i\overline {\epsilon} \overline {\Phi}_{LS}(j))
\end{equation}
\[
U(i,\epsilon)=|U(i)|\exp (i\epsilon \Phi_{US}(i)), \quad 
\overline {U}(j,\overline{\epsilon})=|\overline {U}(j)|\exp (i\overline {\epsilon} \overline {\Phi}_{US}(j))
\]
In the Appendix we show that this four-fold sign ambiguity can be resolved into a single set of phases for each set of moduli $|A(i)|,|\overline {A}(j)|,i,j=1,2$ in the process of determination of the phases of natural exchange amplitudes from the measurements of $Im \rho^0_x$ and $Im \rho^0_z$ on polarized targets.

\section{Amplitude analysis with unpolarized target data on $Re \rho^0_u$.}

We now show that a model independent amplitude analysis is possible with unpolarized target data $Re \rho^0_u$. Perhaps more appropriately, the analysis should be called "intensity analysis". First we define partial wave intensities 
$I_A=|A_u|^2+|A_d|^2=|A|^2+|\overline{A}|^2$, $A=S,L,U,N$. Using the expressions in the Table I we can define a new set of 6 equations 
\begin{eqnarray}
a_1^0 & = &{1 \over{2}}(\rho^0_u)_{SP} \Sigma={1\over{2}}(|I_S|^2+|I_L|^2+|I_U|^2+|I_N|^2)\\
\nonumber
a_2^0 & = &(\rho^0_u)_{PP} \Sigma={1\over{2}}(2|I_L|^2 -|I_U|^2-|I_N|^2)\\
\nonumber
a_3^0 &= &(\rho^0_u)^{11}_{1-1} \Sigma={1\over{2}}(|I_N|^2-|I_U|^2)\\
\nonumber
a_4^0 &= &{1 \over {2}}(\rho^0_u)^{10}_{0s}\Sigma={1\over{2}}|I_L||I_S| \cos(\Psi_{LS})\\
\nonumber
a_5^0 & = &{1 \over{\sqrt{2}}}(\rho^0_u)^{11}_{01} \Sigma={1\over{2}}|I_L||I_U|\cos(\Psi_{LU})\\
\nonumber
a_6^0 &= &{1 \over {\sqrt{2}}}(\rho^0_u)^{10}_{1s} \Sigma={1\over{2}}|I_U||I_S| \cos(\Psi_{US})
\end{eqnarray}
where for $AB=LS,LU,US$ the interference terms are
\begin{equation}
|I_A||I_B| \cos(\Psi_{AB})=Re(AB^*)+Re(\overline{A}\overline{B}^*)
\end{equation}
The system of equations (4.1) can be solved if the correlations $\cos(\Psi_{AB})$ satisfy a cosine condition similar to (3.5). To see if such is the case we used (4.2) to expresse the correlations in terms of transversity amplitudes. Upon substitution into the cosine condition we obtained a non-linear constraint connecting moduli and cosines of relative phases of all transversity amplitudes $S,L,U$ and $\overline{S},\overline{L},\overline{U}$. This condition was tested numerically for all four combinations of the amplitudes $A(i),\overline{A}(j),i,j=1,2$ by every single Monte Carlo sampling of the data error volume that produced a physical solution. In each case the constraint was satisfied exactly as an identity. We concluded that the system (4.1) is analytically solvable using the method described in the previous Section..

To prove the analytical solvability in a simpler way we set in (3.1) and (3.2) all observables $\rho^0_y=0$. Then (4.1) reads
\begin{eqnarray}
a_1^0 & = &{1 \over{2}}(\rho^0_u)_{SP} \Sigma=|S^0|^2+|L^0|^2+|U^0|^2+
|N^0|^2\\
\nonumber
a_2^0 & = &(\rho^0_u)_{PP} \Sigma=2|L^0|^2 -|U^0|^2-|N^0|^2\\
\nonumber
a_3^0 &= &(\rho^0_u)^{11}_{1-1} \Sigma=|N^0|^2-|U^0|^2\\
\nonumber
a_4^0 &= &{1 \over {2}}(\rho^0_u)^{10}_{0s}\Sigma=|L^0||S^0| \cos(\Phi_{LS}^0)\\
\nonumber
a_5^0 & = &{1 \over{\sqrt{2}}}(\rho^0_u)^{11}_{01} \Sigma=|L^0||U^0|\cos(\Phi_{LU}^0)\\
\nonumber
a_6^0 &= &{1 \over {\sqrt{2}}}(\rho^0_u)^{10}_{1s} \Sigma=|U^0||S^0| \cos(\Phi_{US}^0)
\end{eqnarray}
where the amplitudes $A^0=A(\rho^0_u,\rho^0_y=0)$ are analytic continuation of the amplitudes $A=A(\rho^0_u,\rho^0_y)$ to $\rho^0_y=0$. We obtain a similar set of equations from (3.2) with $\overline{a}_k^0, k=1,6$ for amplitudes $\overline{A}^0=\overline{A}(\rho^0_u,\rho^0_y=0)$ which are analytical continuation of the amplitudes $\overline{A}(\rho^0_u,\rho^0_y)$ to $\rho^0_y=0$. Since $\overline{a}_k^0=a_k^0, k=1,6$ we must have
\begin{equation}
A^0=\overline{A}^0 
\end{equation}
The equations (4.1) and (4.3) coincide when
\begin{eqnarray}
I_A=|A|^2+|\overline{A}|^2 & = & 2|A^0|^2=|A^0|^2+|\overline{A}^0|^2=I_A^0\\
\nonumber
\cos(\Psi_{AB}) & = & \cos(\Phi_{AB}^0)
\end{eqnarray}
Since the amplitudes $A^0$ and $\overline{A}^0$ are complex valued functions, they satisfy the phase conditions (3.4) and cosine conditions (3.5). This means that the equations (4.1) are analyticaly solvable.

Helicity nonflip and single flip amplitudes $A_0$ and $A_1$ are related to transversity amplitudes by the relations~\cite{lutz78,svec12b}
\begin{eqnarray}
A_0 & = & {1\over{\sqrt{2}}}(A_u+A_d)={1\over{\sqrt{2}}}(A+\overline{A}\exp(i\omega))\exp(i\Phi(S_u))\\
\nonumber
A_1 & = & {-i\over{\sqrt{2}}}(A_u-A_d)={1\over{\sqrt{2}}}(A-\overline{A}\exp(i\omega))\exp(i\Phi(S_u))
\end{eqnarray}
where $\omega=\Phi(S_d)-\Phi(S_u)$. Then at $\rho^0_y=0$ we have
\begin{eqnarray}
|A_0^0|^2 & = & |A^0|^2+|A^0|^2\cos(\omega)\\
\nonumber
|A_1^0|^2 & = & |A^0|^2-|A^0|^2\cos(\omega)
\end{eqnarray}
In Section VI. we shall find $\cos(\omega)=-1$. Thus $|A_0^0|^2=0$ and $|A_1^0|^2=2|A^0|^2$. We thus find another useful relation for physical intensities in addition to (4.5)
\begin{equation}
I_A=|A|^2+|\overline{A}|^2=|A_1^0|^2=I_A^0
\end{equation}
There are two solutions for the single flip helicity amplitude $A_1^0$ extrapolated to $\rho^0_y=0$ which correspond to the two analytical solutions of the equations (4.1) satisfying the conditions (4.5)
\begin{eqnarray}
|A_1^0(1)|^2 & = & I_A(1,1)=|A(1)|^2+|\overline{A}(1)|^2\\
\nonumber
|A_1^0(2)|^2 & = & I_A(2,2)=|A(2)|^2+|\overline{A}(2)|^2
\end{eqnarray}
The off-diagonal intensities $I_A(1,2)$ and $I_A(2,1)$ are solutions of the equations (4.1) that are not related to the extrapolation of the amplitudes to $\rho^0_y=0$ and thus do not meet the conditions (4.5). The reason for this is the fact that $I_A(1,2) \neq I_A(2,1)$ while $I_A^0(1,2) = I_A^0(2,1)$.

The amplitude analysis determines also the relative phases between single flip amplitudes $A_1^0$. To determine $\Phi(L_1^0)-\Phi(S_1^0)$ we find from (4.6) the interference term
\begin{equation}
L_1S_1^*={1\over{2}}(LS^*+\overline{L}\overline{S}^* -\overline{L}S^*\exp(i\omega)+L\overline{S}^*\exp(-i\omega))
\end{equation}
where $S,\overline{S}$ are real, $L=|L|\exp(i\Phi_{SL})$ and $\overline{L}=|\overline{L}|\exp(i\overline{\Phi}_{SL})$. At $\rho^0_y=0$ we have $S^0=\overline{S}^0$, $L^0=\overline{L}^0$ and $\Phi_{LS}^0=\overline{\Phi}_{LS}^0$. Then (4.10) reads
\begin{equation}
L_1^0S_1^{0*}=|L^0||S^0|\exp(i\Phi_{LS}^0)(1-\cos\omega)
\end{equation}
With $\cos \omega=-1$ we find for both Solutions 1, 2
\begin{equation}
\Phi(L_1^0)-\Phi(S_1^0)=\Phi(L^0_\tau)-\Phi(S^0_\tau)
\end{equation}
where $\tau=u,d$. Note how our results on the single flip amplitudes $A_1^0$ depend on our finding in Section VI. that the relative phases $\Phi(S_d)-\Phi(S_u)=\pm 180^\circ$.

\section{Evidence for $\rho^0(770)-f_0(980)$ mixing: Transversity amplitudes.}

\subsection{Data analysis}

We have performed high resolution amplitude analyses of CERN data on $\pi^- p \to \pi^- \pi^+n$ at 17.2 Gev/c at $0.005 \leq |t| \leq 0.20$ (GeV/c$)^2$ and in dipion mass range $560 \leq m \leq 1080$ MeV with polarized target data $\rho^0_u$, $\rho^0_y$~\cite{chabaud83,rybicki96} (Analysis I) as well as with only unpolarized target data $\rho^0_u$~\cite{grayer74} (Analysis II). The two analyses were performed using the same code but in the Analysis II we set $\rho^0_y=0$. The analysis with polarized target data used 1 million of Monte Carlo samplings of error volumes of measured density matrix elements. Not all values of density matrix elements correspond to physical amplitudes. The procedure produces frequency distribution of physical solutions for the moduli and phases and for various observables calculated from the solutions for amplitudes. The average values of the distributions for the moduli add up exactly to $\Sigma={{d^2 \sigma}/{dtdm}}$ and those for the phases satisfy cosine conditions (3.5). These averaged values can thus be interpreted as the measured amplitudes. The range of distributions determines the asymmetric error bars on the amplitudes and calculated observables. Analysis I using 5 million Monte Carlo samplings produced virtually identical averaged values and slightly larger errors, indicating the analysis is stable. The analysis with unpolarized target data used 200 000 Monte Carlo samplings.

The results from both analyses are shown in Figures 1-7 and 11-13. The Figures show data points with errors from both analyses.

\begin{figure}[htp]
\includegraphics[width=12cm,height=10.5cm]{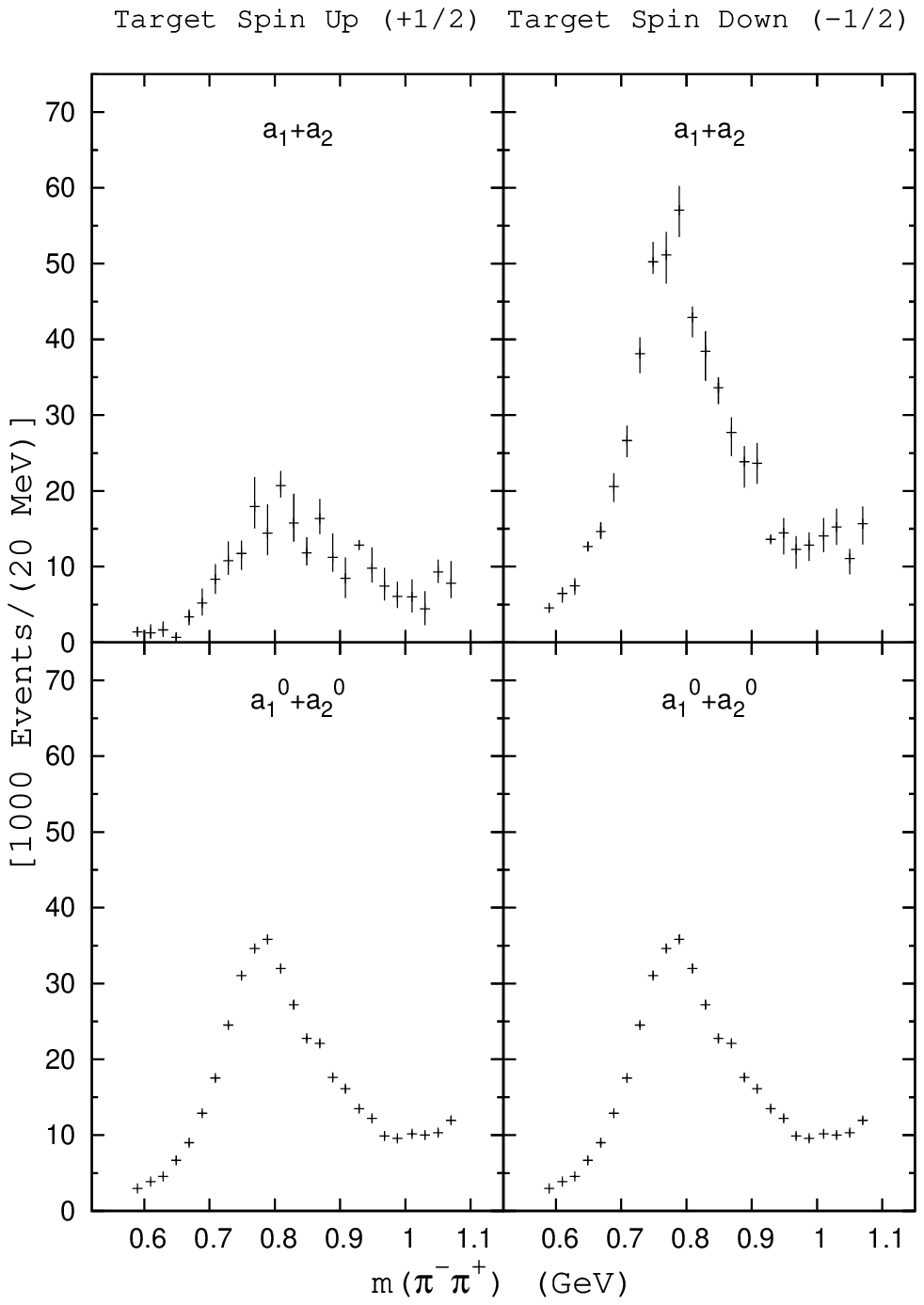}
\caption{Data components $a_1+a_2$ and $a_1^0+a_2^0$ from Analyses I and II, respectively.}
\label{Figure 1.}
\end{figure}

\begin{figure}[hp]
\includegraphics[width=12cm,height=10.5cm]{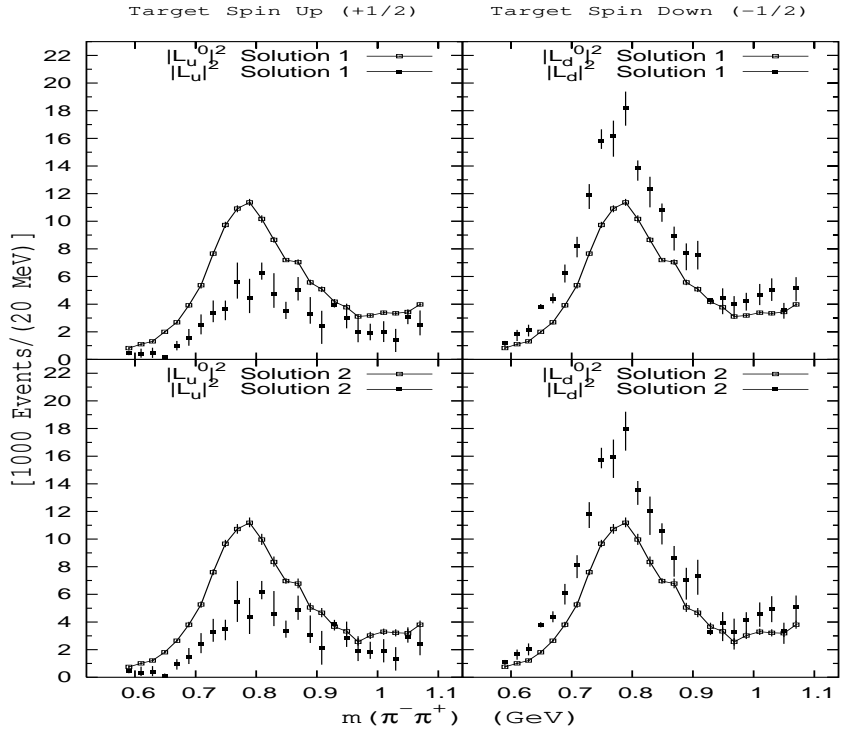}
\caption{Moduli of $P$-wave amplitudes $|L_\tau|^2$ and $|L_\tau^0|^2$ from Analyses I and II (with line).}
\label{Figure 2.}
\end{figure}

\begin{figure}[htp]
\includegraphics[width=12cm,height=10.5cm]{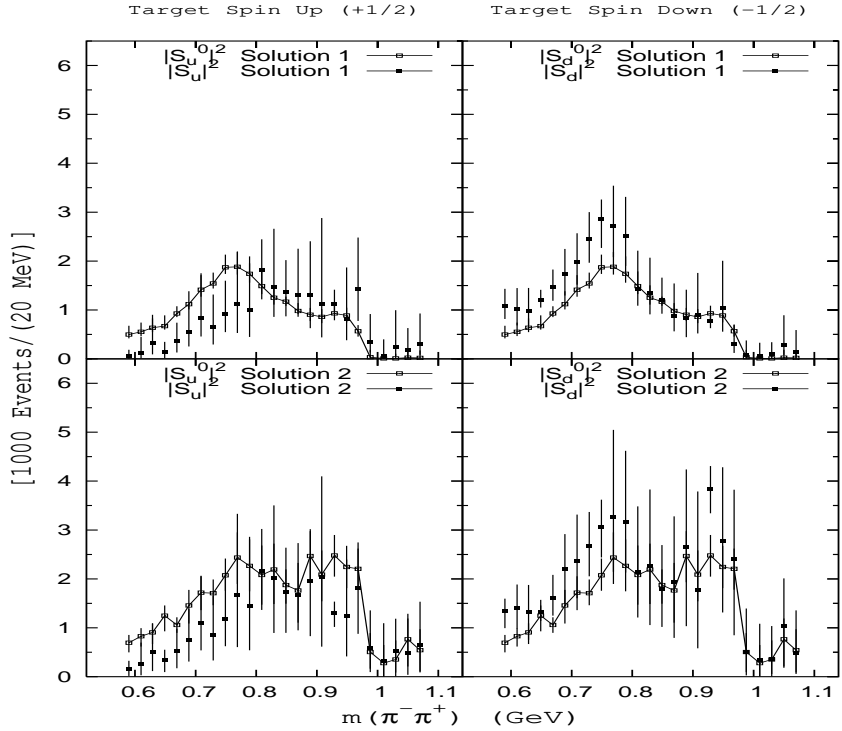}
\caption{Moduli of $S$-wave amplitudes $|S_\tau|^2$ and $|S_\tau^0|^2$ from Analyses I and II (with line).}
\label{Figure 3.}
\end{figure}

\subsection{Evidence from the moduli}

To understand the origin of the $\rho^0(770)-f_0(980)$ mixing in the $S$-wave amplitudes we recall from (3.7) the relation
\begin{equation}
|S|^2 = (a_1+a_2)-3|L|^2
\end{equation}
 in the Analysis I and a similar equation for the amplitudes $|S^0|^2$ in terms of $a_1^0+a_2^0$ in the Analysis II. We show that the presence of $\rho^0(770)$ in the $S$-wave amplitudes in both analyses arises from the dominant presence of this resonance in the data components $a_1+a_2$ and $a_1^0+a_2^0$. The data components were calculated from those Monte Carlo samplings of density matrix elements in the error volume of the data for which physical solutions for the amplitudes were found. The results for $a_1+a_2$ and $a_1^0+a_2^0$ are shown in Figure 1. The Figure shows pronounced $\rho^0(770)$ peak for target spin "down" component and a suppression of $\rho^0(770)$ in the target spin "up" component. There are pronounced dips at 970 MeV and 1030 MeV in target spin "down" and "up" components, respectively, corresponding to the $f_0(980)$ resonance. In contrast, the components $a^0_1+a_2^0$ are equal for both target transversities, peak at $\rho^0(770)$ mass and dip at 980 MeV.

\begin{figure}[hp]
\includegraphics[width=12cm,height=10.5cm]{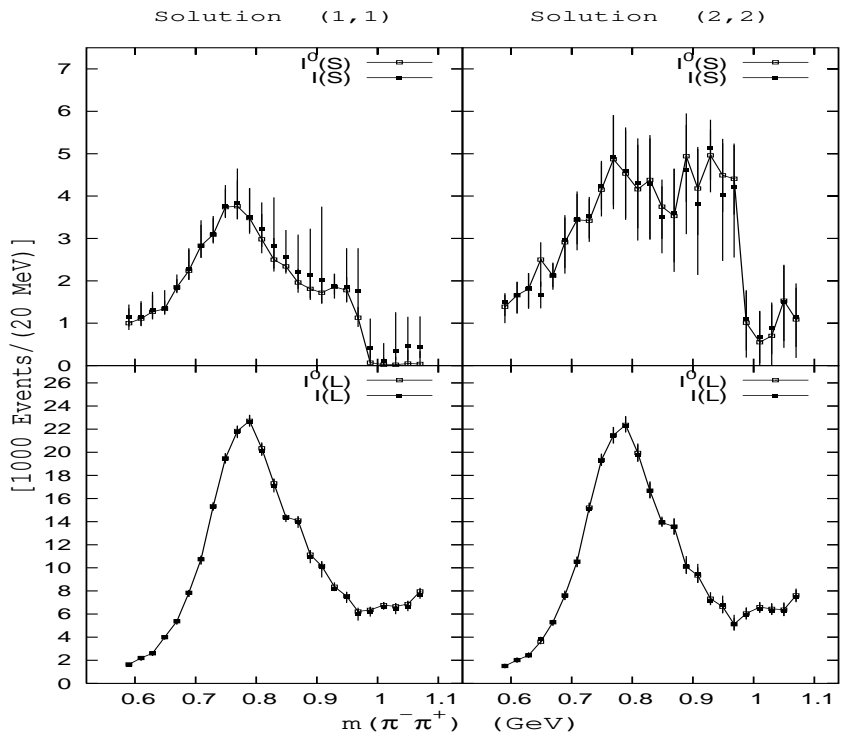}
\caption{Intensities $I_A$ and $I_A^0$ for $S$- and $P$-wave amplitudes from Analyses I and II (with line).}
\label{Figure 4.}
\end{figure}

Figure 2 shows the two solutions for the $P$-wave amplitudes $|L_u|^2$ and $|L_d|^2$ from the Analysis I and the amplitudes $|L_u^0|^2=|L_d^0|^2$ from the Analysis II. Remarkably, in both analyses the two solutions of the two cubic equations reproduce  closely the general features of the data components in the Figure 1. In the Analysis I the $\rho^0(770)$ peak is enhanced in $|L_d|^2$ and suppressed in $|L_u|^2$. 

Figure 3 shows the two solutions for the $S$-wave amplitudes $|S_u|^2$ and 
$|S_d|^2$ from the Analysis I and amplitudes $|S_u^0|^2=|S_d^0|^2$ from the Analysis II. The rho-like resonance and $f_0(980)$ are clearly resolved in both solutions for $|S_d|^2$ in Analysis I. This pronounced peak at 770 MeV with a width at half-height of $\sim 155$ MeV has its origin in the $\rho^0(770)$ peak of the data component $a_1+a_2$ which peak survives the subtraction of $3|L_d|^2$. In the Analysis I the $\rho^0(770)$ peak is enhanced in $|S_d|^2$ and suppressed in $|S_u|^2$. The Analysis II shows a clear $\rho^0(770)$ peak in Solution 1 and a subdued $\rho^0(770)$ peak in Solution 2 in the amplitude 
$|S_d^0|^2$. 

In both analyses the amplitudes $|S_d|^2$ and $|S_d^0|^2$ show a shoulder at $\sim 930$ MeV in Solution 1 and a secondary peak at the same mass in Solution 2. In Ref.~\cite{svec02a} we have shown in fits to the moduli of $S$-wave amplitudes that these structures arise from the interference of $\rho^0(770)$ and $f_0(980)$ Breit-Wigner amplitudes with a complex backgroud.

In Figure 4 we present the partial wave intensities $I_S(i,j),I_L(i,j)$ from the Analysis I and $I^0_S(i,j),I_L^0(i,j)$ from the Analysis II for the Solutions $(i,j)=(1,1), (2,2)$. The intensities from the two analyses are virtually identical, although they come from two different calculations with differenet sets and numbers of physical Monte Carlo samplings. This result confirms the validity of our analysis with unpolarized target data. Both analyses show the expected dominant $\rho^0(770)$ peak in the Solutions $I_S(1,1)$ and $I_S^0(1,1)$. The crucial result however is the observation that both analyses show a clear albeit subdued $\rho^0(770)$ peak in the Solutions $I_S(2,2)$ and $I_S^0(2,2)$. This means that an exact amplitude analysis of unpolarized target data provides evidence for the presence of $\rho^0(770)$ in both single flip amplitudes $|S_1^0(1)|^2$ and $|S_1^0(2)|^2$.

The observation of $\rho^0(770)-f_0(980)$ mixing in the $S$-wave amplitudes brings up the question of $\rho^0(770)-f_0(980)$ mixing in the $P$-wave amplitudes. Figure 2 shows a clear dip at 970 MeV in the amplitude $|L_d|^2$ followed by a sharp rise. The dip occurs at the mass of scalar resonance $f_0(980)$ and is more pronounced in the Solution 2. This finding is a strong indication for $\rho^0(770)-f_0(980)$ mixing in the $P$-wave amplitudes $L_\tau$. Since $f_0(980)$ resonance is produced by unnatural exchange we expect $f_0(980)$ to be suppressed in natural exchange amplitudes $|N_\tau|^2$ as evidenced by the data in Figure 12.

\subsection{Evidence from the relative phases}

Figures 5, 6 and 7 show the mass dependence of the phases $\Phi_{AB}$ and $\overline {\Phi}_{AB}$, $AB=LS,US,LU$ from the Analysis I. As the figures show, these phases are continous and nearly constant functions. From this fact we can understand why the phases $\Phi_{LS}, \overline {\Phi}_{LS}$ cannot change sign. From (3.16) we see that a change of signs of $\Phi_{LS}$ or $ \overline {\Phi}_{LS}$ results in the change of sign of all phases of the same transversity. This would result in large unphysical discontinuities in the phases $\Phi_{US}, \overline {\Phi}_{US}$ and $\Phi_{LU}, \overline {\Phi}_{LU}$. 

\begin{figure}[htp]
\includegraphics[width=12cm,height=10.5cm]{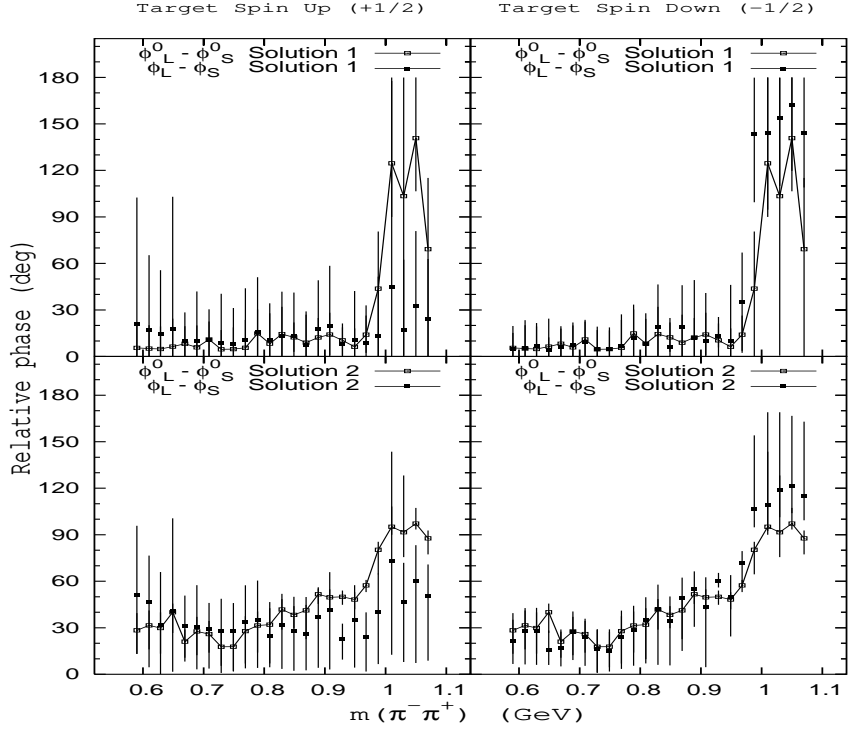}
\caption{Relative phases $\Phi_{LS}=\Phi_{L_u}-\Phi_{S_u}$ and 
$\overline {\Phi}_{LS}=\Phi_{L_d}-\Phi_{S_d}$ from Analyses I and II (with line).}
\label{Figure 5.}
\end{figure}

\begin{figure}[hp] 
\includegraphics[width=12cm,height=10.5cm]{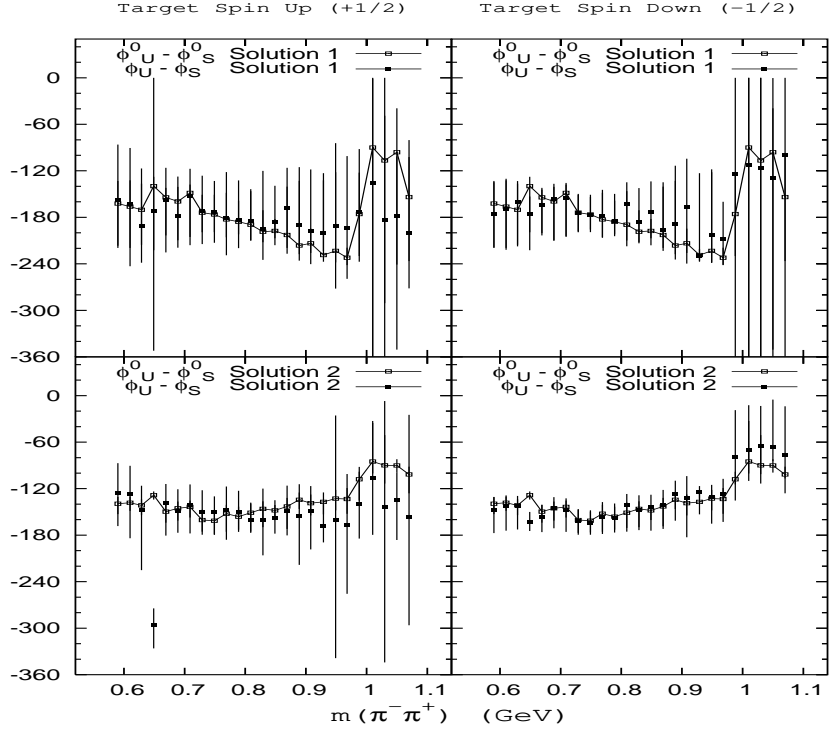}
\caption{Relative phases $\Phi_{US}=\Phi_{U_u}-\Phi_{S_u}$ and 
$\overline {\Phi}_{US}=\Phi_{U_d}-\Phi_{S_d}$ from Analyses I and II (with line).}
\label{Figure 6.}
\end{figure}

\begin{figure}[t]
\includegraphics[width=12cm,height=10.5cm]{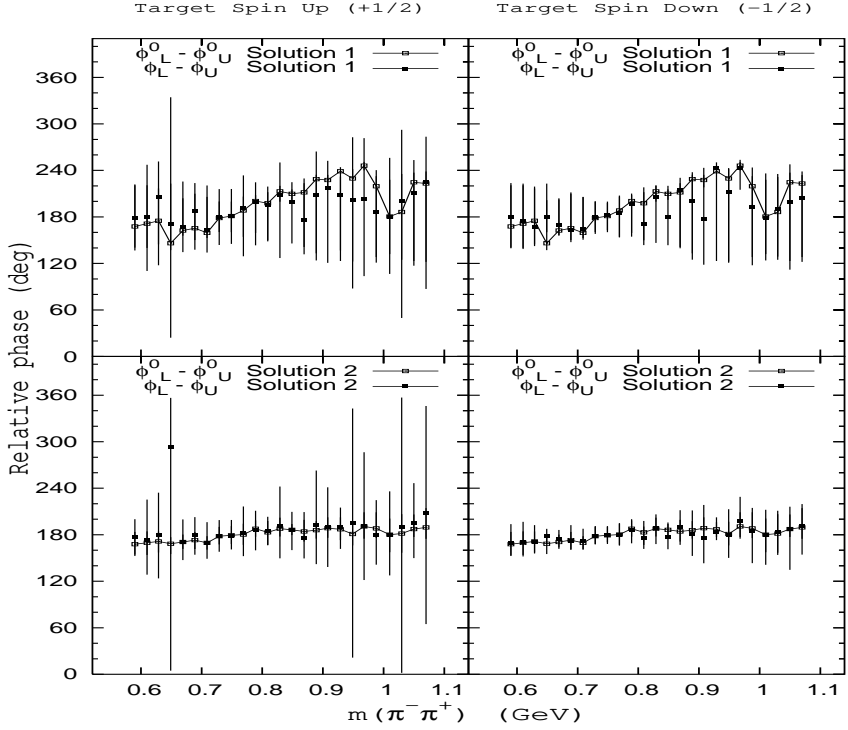}
\caption{Relative phases $\Phi_{LU}=\Phi_{L_u}-\Phi_{U_u}$ and 
$\overline {\Phi}_{LU}=\Phi_{L_d}-\Phi_{U_d}$ from Analyses I and II (with line).}
\label{Figure 7.}
\end{figure}

Figures 5, 6 and 7 also show the mass dependence of the phases $\Phi_{AB}^0$, $\overline {\Phi}_{AB}^0$, $AB=LS,US,LU$ from the Analysis II. There is a remarkable agreement between the Analyses I and II except for Solution 1 above 900 MeV in phases  $\Phi_{LS}^0$, $\overline {\Phi}_{LS}^0$, $\Phi_{US}^0, \overline {\Phi}_{US}^0$. While the phases $\Phi_{AB}$ and $\overline{\Phi}_{AB}$ are similar, the phases $\Phi_{AB}^0$ and $\overline{\Phi}_{AB}^0$ are equal. Due to the similarity of phases $\Phi_{LS}$ and $\overline{\Phi}_{LS}$ we may conclude that not only 
$\Phi(L_1^0)-\Phi(S_1^0)=\Phi(L^0_\tau)-\Phi(S^0_\tau), \tau=u,d$ but also
$\Phi(L_1^0)-\Phi(S_1^0) \approx \Phi(L_\tau)-\Phi(S_\tau)$ for both Solutions 1 and 2.

Independent evidence for the presence of $\rho^0(770)$ in the $S$-wave amplitudes comes from the relative phases $\Phi_{LS}= \Phi_{L_u}-\Phi_{S_u}$ and $\overline {\Phi}_{LS}= \Phi_{L_d}-\Phi_{S_d}$ from both analyses. For target spin "up" the phase $\Phi_{LS}$ is constant and small, reflecting the similarity of $|S_u|^2$ and $|L_u|^2$ in both Solutions 1 and 2. For target spin "down" the phase $\overline {\Phi}_{LS}$ is near zero in the Solution 1 and small and slowly varying in Solution 2 below 1000 MeV. These results indicate that the amplitudes $S_\tau$ and $L_\tau$ are nearly in phase. Since the amplitudes $L_\tau$ resonate at $\rho^0(770)$, so must the amplitudes $S_\tau$. 

This conclusion is in agreement with the phases $\Phi_{US}= \Phi_{U_u}-\Phi_{S_u}$, $\overline {\Phi}_{US}= \Phi_{U_d}-\Phi_{S_d}$ from both analyses. These phases are approximately constant below 1000 MeV and indicate that the amplitudes $S_\tau$ and $U_\tau$ are nearly 180$^\circ$ out of phase. Since amplitudes $U_\tau$ resonate at $\rho^0(770)$, so must the amplitudes $S_\tau$. The Figure 7 shows that the relative phases of the amplitudes $L_\tau$ and $U_\tau$ are nearly constant and nearly 180$^\circ$ out of phase, as expected from these two $P$-wave amplitudes.

\section{Evidence for $\rho^0(770)-f_0(980)$ mixing: Helicity amplitudes.}

\subsection{Helicity amplitudes}

Amplitude analysis of the complete $S$- and $P$-wave subsystem of the reduced density matrix yields four sets of solutions for reduced transversity amplitudes $A(i),\overline{A}(j), i,j=1,2$. For unnatural exchange amplitudes $A=S,L,U$ the reduced transversity amplitudes are related to transversity amplitudes by phase factors
\begin{eqnarray}
A_u(i) & = & A(i) \exp {i \Phi_{S_u}(i)}\\
\nonumber
A_d(j) & = & \overline {A}(j) \exp {i \omega_{ij}} \exp {i \Phi_{S_u}(i)}
\end{eqnarray}
where  $\Phi_{S_u}(i)$ is the arbitrary absolute phase and
\begin{equation}
\omega_{ij}=\Phi_{S_d}(j)-\Phi_{S_u}(i)
\end{equation}
is the relative phase between $S$-wave amplitudes of opposite transversity. Recall from (2.6) that $S(i),\overline{S}(j)$ are real and positive amplitudes, the phases of $L(i)$ and $\overline{L}(j)$ are $\Phi_{LS}(i)$ and $\overline{\Phi}_{LS}(j)$, and the phases of $U(i)$ and $\overline{U}(j)$ are $\Phi_{US}(i)$ and $\overline{\Phi}_{US}(j)$, respectively.

Helicity amplitudes $A^J_{\lambda \chi,0 \nu}$ with definite $t$-channel naturality were defined and related to transversity amplitudes of definite $t$-channel naturality in Ref.~\cite{lutz78,svec12b} for any dipion spin $J$ and helicity $\lambda$. Due to the $P$-parity conservation only helicity nonflip and helicity flip amplitudes $A^J_{\lambda,0}=A^J_{\lambda +,0 +}$ and $A^J_{\lambda,1}=A^J_{\lambda +,0 -}$ are independent. Here $n=0,1$ is nucleon helicity flip $n=|\chi-\nu|$. The $S$- and $P$-wave helicity amplitudes $A_n$ are related to the transversity amplitudes $A_\tau$ by 
relations~\cite{lutz78,svec12b}
\begin{equation}
A_n={(-i)^n \over{\sqrt{2}}}(A_u+(-1)^n A_d)
\end{equation}
In terms of reduced transversity amplitudes we can write for the unnatural exchange amplitudes
\begin{equation}
A_n(ij)={(-i)^n \over{\sqrt{2}}}(A(i)+(-1)^n \overline{A}(j)
\exp(i \omega_{ij}))\exp(i\Phi_{S_u}(i))
\end{equation}
Each set of solutions comes with a fourfold sign ambiguity in the phases of the amplitudes $(A(i),\overline{A}(j))_{\epsilon \overline{\epsilon}}$ where 
$\epsilon \overline{\epsilon}=++,+-,-+,--$ are the signs of relative phases $\Phi_{LS}(i)$ and $\overline{\Phi}_{LS}(j)$. From (3.16) we see that the change of sign of $\Phi_{LS}(i)$ ($\overline{\Phi}_{LS}(j)$) results in the change of sign of the phase of amplitude $U(i)$ ($\overline{U}(j)$), or complex conjugation of amplitudes $A(i)$ ($\overline{A}(j)$). We can thus write the four sets of phases for each $i,j$
\begin{eqnarray}
(A(i),\overline{A}(j))_{++} & = & (A(i),\overline{A}(j))\\
\nonumber
(A(i),\overline{A}(j))_{+-} & = & (A(i),\overline{A}(j)^*)\\
\nonumber
(A(i),\overline{A}(j))_{-+} & = & (A(i)^*,\overline{A}(j))=((A(i),\overline{A}(j))_{+-})^*\\
\nonumber
(A(i),\overline{A}(j))_{--} & = & (A(i)^*,\overline{A}(j)^*)=((A(i),\overline{A}(j))_{++})^*
\end{eqnarray}
As shown in the Appendix, the measurements with planar target polarization can select uniquely one of the four solutions for the phases. Since the transversity amplitudes $A_u(i)$ and $A_d(j)$ are also complex conjugated, we get for helicity amplitudes
\begin{eqnarray}
A_n(ij)_{-+} & = & (-)^n(A_n(ij)_{+-})^*\\ 
\nonumber
A_n(ij)_{--} & = & (-)^n(A_n(ij)_{++})^*
\end{eqnarray}
Since we do not get new solutions with the phase sets $--$ and $-+$ numerical calculations were done only for the sets with phases $++$ and $+-$.

\subsection{Bilinear terms of the helicity amplitudes}

We now look at the bilinear terms $A_nA_n^*=|A_n|^2$, $A=S,L,U$ and $A_nB_n^*$,$AB=LS,US,UL$.
For the sake of brevity we shall omit in the following the indices $ij$ and $++$, $+-$. Using (6.4) or (6.6) we obtain
\begin{eqnarray}
|A_n|^2 & = & {1\over{2}} \Bigl (|A|^2+|\overline{A}|^2+(-1)^n2X_A \cos(\omega)+
(-1)^n2Y_A \sin(\omega) \Bigr )\\
\nonumber
& = & {1\over{2}} \Bigl (I_A+(-1)^n2X_A \cos(\omega)+(-1)^n2Y_A \sin(\omega) \Bigr )
\end{eqnarray}
where $X_A=Re(A \overline{A}^*)$, $Y_A=Im(A \overline{A}^*)$ and $I_A=|A|^2+|\overline{A}|^2$
is partial wave intensity. Note that $Y_S=Im(S \overline{S}^*)=0$ as both $S$ and $\overline{S}$ are real. For the bilinear terms $A_nB_n^*$ we obtain
\begin{equation}
A_nB_n^*={1\over{2}} \Bigl ( AB^* + \overline{A}\overline{B}^* +(-1)^n \bigl (  A \overline{B}^* e^{-i\omega} +\overline{A}B^* e^{+i\omega} \bigr ) \Bigr )
\end{equation}
The real part reads
\begin{equation}
Re(A_n B_n^*)={1\over{2}} \Bigl ( Re(AB^*)+Re(\overline{A} \overline{B}^*)
\end{equation}
\[
+(-1)^n \bigl ( (Re(A\overline{B}^*)+Re(\overline{A}B^*) ) \cos\omega
+(Im(A \overline{B}^*)-Im(\overline{A} B^*) )\sin \omega \bigr ) \Bigr )
\]
The imaginary part reads
\begin{equation}
Im(A_n B_n^*)={1\over{2}} \Bigl ( Im(AB^*)+Im(\overline{A} \overline{B}^*)
\end{equation}
\[
+(-1)^n \bigl ( (Im(A\overline{B}^*)+Im(\overline{A}B^*) ) \cos\omega
-(Re(A \overline{B}^*)-Re(\overline{A} B^*) )\sin \omega \bigr ) \Bigr )
\]

It is apparent from (6.4) and (6.6) that the knowledge of $\omega$ allows to determine the helicity amplitudes up to an absolute phase for each solution set $i,j$ and phase set $++$ and $+-$. As we show in the next Section, the phase $\omega$ can be determined analytically from the consistency condition
\begin{equation}
|A_n|^2|B_n|^2=(Re(A_nB_n^*))^2+(Im(A_nB_n^*))^2
\end{equation}
where the terms are given by (6.7),(6.9) and (6.10).

\subsection{Analytical solutions of the relative phase $\omega$}

In order to make use of the consistency condition (6.11) to determine $\omega$, we first rewrite the terms (6.8), (6.10) and (6.11) in a simplified form. To this end we define
\begin{eqnarray}
X(AB) & = & Re(AB^*)+Re(\overline{A} \overline{B}^*)\\
\nonumber
Y(AB) & = & Im(AB^*)+Im(\overline{A} \overline{B}^*)\\
\nonumber
X(A\overline{B})_+ & = & Re(A\overline{B}^*)+Re(\overline{A}B^*)\\
\nonumber
Y(A\overline{B})_+ & = & Im(A\overline{B}^*)+Im(\overline{A}B^*)\\
\nonumber
X(A\overline{B})_- & = & Re(A\overline{B}^*)-Re(\overline{A}B^*)\\
\nonumber
Y(A\overline{B})_- & = & Im(A\overline{B}^*)-Im(\overline{A}B^*)
\end{eqnarray}
Then
\begin{eqnarray}
Re(A_nB_n^*) & = & {1\over{2}} \Bigl ( X(AB)+(-1)^n \bigl (X(A\overline{B})_+ \cos \omega +Y(A\overline{B})_- \sin \omega \bigr ) \Bigr )\\
\nonumber
Im(A_nB_n^*) & = & {1\over{2}} \Bigl ( Y(AB)+(-1)^n \bigl (Y(A\overline{B})_+ \cos \omega -X(A\overline{B})_- \sin \omega \bigr ) \Bigr )\\
\nonumber
     |A_n|^2 & = & {1\over{2}} \Bigl ( X(AA)+(-1)^n \bigl (X(A\overline{A})_+  
\cos \omega +Y(A\overline{A})_- \sin \omega \bigr ) \Bigr )
\end{eqnarray}
Note that $X(AA)=I_A$, $X(A\overline{A})_+=2X_A$ and $Y(A\overline{A})_-=2Y_A$. We can write
\begin{eqnarray}
Re(A_nB_n^*) & = & {1\over{2}} \Bigl ( X(AB)+(-1)^n XG(A \overline{B}) \Bigr )\\
\nonumber
Im(A_nB_n^*) & = & {1\over{2}} \Bigl ( Y(AB)+(-1)^n YG(A \overline{B}) \Bigr )\\
\nonumber
     |A_n|^2 & = & {1\over{2}} \Bigl ( X(AA)+(-1)^n XG(A \overline{A}) \Bigr )\\
\nonumber
     |B_n|^2 & = & {1\over{2}} \Bigl ( X(BB)+(-1)^n XG(B \overline{B}) \Bigr )
\end{eqnarray}
where
\begin{eqnarray}
XG(A \overline{B}) & = & X(A\overline{B})_+ \cos \omega +Y(A\overline{B})_- \sin \omega\\
\nonumber
YG(A \overline{B}) & = & Y(A\overline{B})_+ \cos \omega -X(A\overline{B})_- \sin \omega 
\end{eqnarray}
Next we require that
\begin{equation}
|A_n|^2|B_n|^2=(Re(A_nB_n^*))^2+(Im(A_nB_n^*))^2
\end{equation}
The l.h.s. of (6.16) reads
\begin{equation}
X(AA)X(BB)+XG(A\overline{A})XG(B\overline{B}) 
\end{equation}
\[
+(-1)^n X(AA)XG(A\overline{A})X+(-1)^nX(BB)XG(A\overline{A})
\]
The r.h.s. of (6.16) reads 
\begin{equation}
X(AB)^2+XG(A \overline{B})^2+Y(AB)^2+YG(A \overline{B})^2
\end{equation}
\[
+(-1)^n2X(AB)XG(A \overline{B})+(-1)^n2Y(AB)YG(A \overline{B})
\]
Subtracting (6.16) with $n=1$ from (6.16) with $n=0$ and using (6.15) we obtain equation linear in $\cos \omega$ and $\sin \omega$
\begin{equation}
\sin \omega \Bigl ( X(AA)Y(B\overline{B})_- +X(BB)Y(A\overline{A})_- -2X(AB)Y(A\overline{B})_- +2Y(AB)X(A\overline{B})_- \Bigr ) =
\end{equation}
\[
-\cos \omega \Bigl ( X(AA)X(B\overline{B})_+ +X(BB)X(A\overline{A})_+ -2X(AB)X(A\overline{B})_+ -2Y(AB)Y(A\overline{B})_+ \Bigr )
\]
which can be cast in the form
\begin{equation}
\sin \omega W_2 = - \cos \omega W_1
\end{equation}
Using $\sin^2 \omega + \cos^2 \omega =1$ we find
\begin{eqnarray}
\cos \omega & = & {\pm W_2 \over{W}}\\
\nonumber
\sin \omega & = & {\mp W_1 \over{W}}
\end{eqnarray}
where $W=\sqrt{W_1^2+W_2^2}$.

Using (6.7), (6.9) and (6.10) it is straightforward to verify that the consistency condition (6.16) reduces to identity in the following two cases
\begin{eqnarray}
\cos \omega =0, & \sin \omega = \pm 1\\
\sin \omega =0, & \cos \omega = \pm 1
\end{eqnarray}
The first case leads to two solutions $\omega = + \pi/2$ and $\omega= -\pi/2$. The second case leads to another two solutions $\omega=0$ and $\omega=\pi$.

\begin{figure} 
\includegraphics[width=12cm,height=10.5cm]{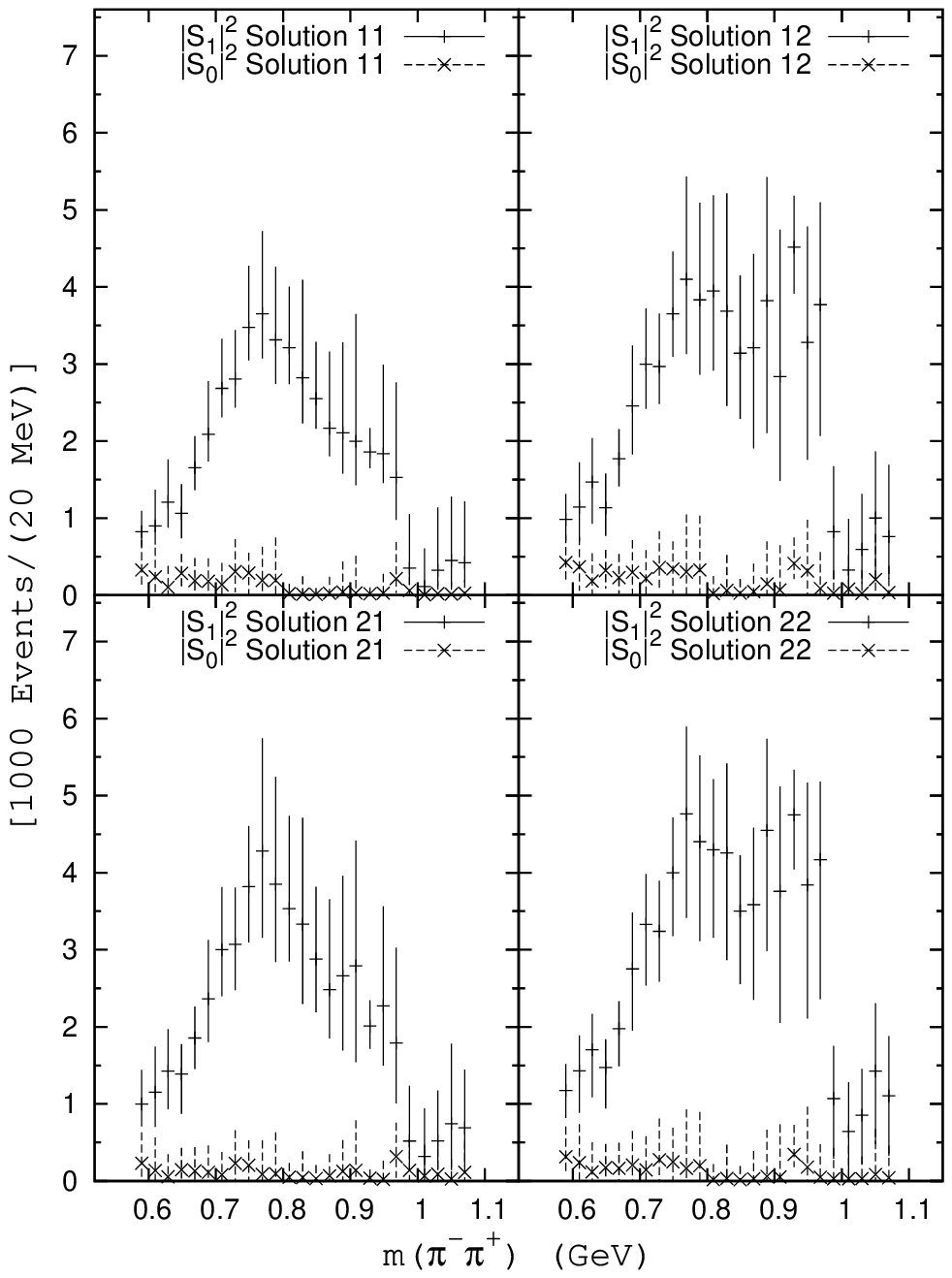}
\caption{Moduli $|S_0(ij)|^2$ and $|S_1(ij)|^2$ for $i,j=1,2$. The solutions for the phase signs ++ and +-are equal.}
\label{Figure 8}
\end{figure}

\begin{figure} 
\includegraphics[width=12cm,height=10.5cm]{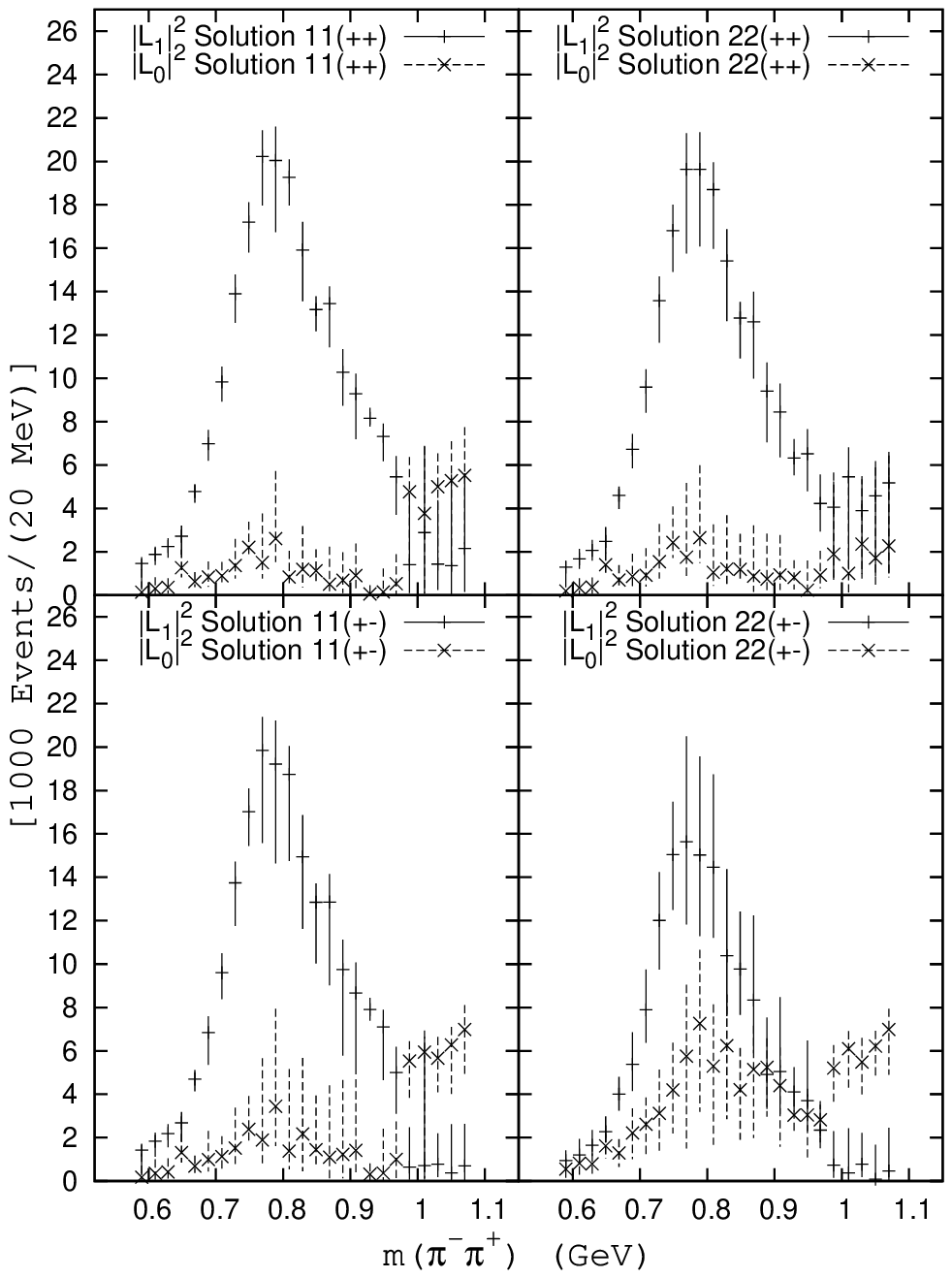}
\caption{Moduli $|L_0(ij)|^2$ and $|L_1(ij)|^2$ for $ij=11,22$. The solutions for the phase signs ++ and +- are different.}
\label{Figure 9}
\end{figure}

\begin{figure} 
\includegraphics[width=12cm,height=10.5cm]{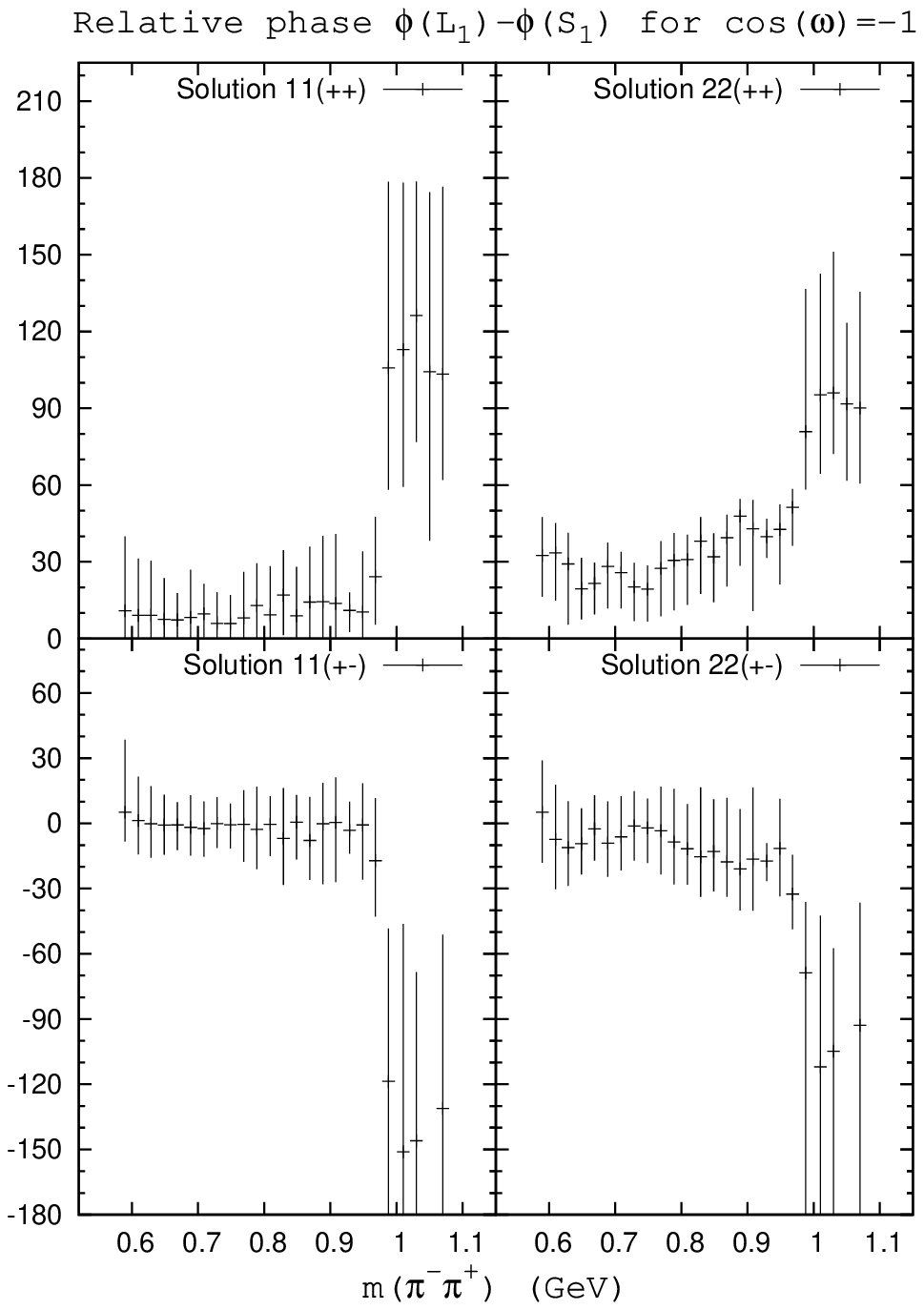}
\caption{Relative phase $\Phi(L_1)-\Phi(S_1)$ for $ij=11,22$. The solutions for the phase signs ++ and +- are different.}
\label{Figure 10}
\end{figure}

\subsection{Solutions for $S$- and $P$-wave helicity amplitudes}

\subsubsection{Numerical calculations and their checks}

The Monte Carlo amplitude analysis of reduced transversity amplitudes and related observables was carried out by a computer code A. The mean values of moduli and phases of the reduced transversity amplitudes satisfy strict normalization and phase conditions, respectively, and thus represent the true measured amplitudes at the $t$-bin average of 0.067 (GeV/c)$^2$. These mean values were used as an input in an exploratory computer code B to calculate $\omega$ using (6.21) and helicity amplitudes $A_n,n=0,1$, $A=S,L,U$ using $\omega$ from (6.21), (6.22) and (6.23). Errors on $\omega$ from (6.21) and on helicity amplitudes were not calculated due to high non-linearity of the equations.

Equations (6.21) were used with amplitudes $A=L,B=S$ to calculate $\omega(ij)$ for each set $++$ and $+-$ of signs of phases. Using these values of $\omega(ij)$ as well as the values corresponding to the solutions (6.22) and (6.23), moduli $|A_n|^2,|B_n|^2$ and interference terms $Re(A_nB_n^*),Im(A_nB_n^*)$ were calculated to determine cosines and sines of the relative phases 
\begin{equation}
\Phi(A_nB_n^*)=\Phi_{A_n}-\Phi_{B_n}
\end{equation}
Since the calculations of $\omega$, moduli and the interference terms are all entirely independent, the selfconsistency of the data and calculations was checked using the following three tests on the relative phases. The first test are trigonometric identities
\begin{equation}
\cos^2 \Phi(A_nB_n^*) + \sin^2 \Phi(A_nB_n^*)=1
\end{equation}
The second test are phase conditions
\begin{equation}
(\Phi_{L_n}-\Phi_{S_n})-(\Phi_{U_n}-\Phi_{S_n})+(\Phi_{U_n}-\Phi_{L_n})=0
\end{equation}
The third test are cosine conditions equivalent to phase conditions
\begin{equation}
\cos^2 \Phi(L_nS_n^*)+\cos^2 \Phi(U_nS_n^*)+\cos^2 \Phi(U_nL_n^*)-
2\cos \Phi(L_nS_n^*)\cos \Phi(U_nS_n^*)\cos \Phi(U_nL_n^*)=1
\end{equation}
All conditions are satisfied identically within the single precision calculation used by the code for all combinations of solutions $i,j=1,2$ for both sets $++$ and $+-$ of signs of phases. For instance, typical deviation from 1 of the trigonometric identities (6.25) are of order $10^{-5} - 10^{-7}$.

To select a physical solution we used two criteria:
(1) The single flip amplitudes must dominate non-flip amplitudes due to the pion exchange dominance at small $t$.
(2) The amplitudes $L_1$ and $U_1$ must show a clear $\rho^0(770)$ peak. 

The computer codes A and B were merged in a computer code C to calulate errors on helicity amplitudes using a physical solution for $\omega$ found in the exploratory code B. Each Monte Carlo selection from the data error volume that yields a physical solution of transversity amplitudes also yields the corresponding physical solution of helicity amplitudes, enabling code C to calculate their average value and error range in each mass bin in the same way as for the transversity amplitudes. The transversity and helicity amplitudes with errors thus represent identical experimental data. 

\subsubsection{Unphysical solutions for helicity amplitudes}

The values of $\omega(ij)_{++}$ and $\omega(ij)_{+-}$ calculated from (6.21) show random variations as a function of dipion mass for all solution sets $i,j=1,2$ and signs of phases. All amplitudes exhibit the same random behaviour and do not show the required resonant Breit-Wigner behaviour at $\rho^0(770)$ resonance. Moreover, the non-flip amplitude $L_0$ has magnitude comparable to or larger than the flip amplitude $L_1$. These solutions with $\cos \omega \neq 0$ and $\sin \omega \neq 0$ are rejected as unphysical.

The two values for $\sin \omega = \pm 1$ from (6.22) correspond to $\omega_{ij,++}=\pm \pi/2$ and $\omega_{ij,+-}= \pm \pi/2$ for all solution sets $i,j=1,2$. The solutions with $\sin \omega =+1$ are excluded because they require that the non-flip amplitude $L_0$ is larger than the flip amplitude $L_1$. Because $\cos \omega=0$ and the amplitudes $S_0$ and $S_1$ do not depend on $\sin \omega$, both solutions $\sin \omega =\pm 1$ give $|S_0|^2=|S_1|^2=I_S/2$ where $I_S$ is $S$-wave intensity. As a result, both values $\sin \omega = \pm 1$ are unphysical for any solution set $i,j=1,2$ and signs of phases $++$ and $+-$. 

The solutions with $\cos \omega=+1$ ($\omega=0$) from (6.23) require that 
$|S_1|^2 <|S_0|^2$ and $|L_1|^2<|L_0|^2$ for any solution set $i,j=1,2$ and signs of phases. The magnitudes of $S_1$ are small compared to magnitudes of $S_0$, in contradiction with the pion exchange dominance. These solutions are thus excluded as unphysical solutions.

\subsubsection{Physical solutions for amplitudes $S_n$ and $L_n$}

In the solutions with $\cos \omega=-1$ ($\omega=\pm 180^\circ$) the pion exchange dominance is observed in all amplitudes and $|L_1|^2$ and $|U_1|^2$ show a well defined $\rho^0(770)$ peak. The resulting amplitudes thus represent a unique solution to both helicity and transversity amplitudes, up to signs of the phases $\Phi_{LS}$ and $\overline{\Phi}_{LS}$ to be resolved by the measurements of $Im \rho^0_x$ and $Im \rho^0_y$. The solution $\cos \omega=-1$ was used in the computer code C to calulate average values and errors of helicity amplitudes in a Monte Carlo analysis of the data.

Figure 8 shows the moduli of $S$-wave helicity amplitudes. The solutions for the phase signs $++$ and $+-$ are equal. All solutions for 
$|S_1(ij)|^2,i,j=1,2$ provide evidence for $\rho^0(770)-f_0(980)$ mixing in the $S$-wave helicity flip amplitudes. The $\rho^0(770)$ peak stands out in in Solutions $|S_1(11)|^2$ and $|S_1(21)|^2$ which resemble the Solution 1 of $|S_d|^2$. The peak is clearly visible in the Solutions $|S_1(12)|^2$ and $|S_1(22)|^2$ which are similar to Solution 2 of $|S_d|^2$. The secondary structure in the region of 900-1000 MeV is due to $f_0(980)$.

Figure 9 shows the moduli of helicity amplitudes $L_0$ and $L_1$. The Solutions $(2,1)$ and $(1,2)$ (not shown) are similar to Solutions $(1,1)$ and $(2,2)$, respectively. The helicity flip amplitudes $|L_1|^2$ dominate the small non-flip amlitudes with the $\rho^0(770)$ peak below 980 MeV. Above 980 MeV $|L_1|^2$ decreases rapidly while there is a sudden rise in 
$|L_0|^2$. These structures are associated with $f_0(980)$ resonance and support the suggestion of $\rho^0(779)-f_0(980)$ mixing in the helicity non-flip amplitude $L_0$. The exception is the Solution $(2,2)++$ where the rise is observed in $|L_1|^2$.

Figure 10 shows the relative phase $\Phi(L_1)-\Phi(S_1)$ which provide an independent evidence for $\rho^0(779)-f_0(980)$ mixing in the $S$-wave. Again, the Solutions $(2,1)$ and $(1,2)$ (not shown) are similar to Solutions $(1,1)$ and $(2,2)$, respectively. In all solutions the phase is small and nearly constant below 960 MeV. Since $L_1$ resonates near $\rho^0(770)$ with a Breit-Wigner phase $\Phi(\rho^0)$, the phase of the amplitude $S_1$ must be nearly the same resonant phase. The sudden rise of the phase $\Phi(L_1)-\Phi(S_1)$ in the $f_0(980)$ mass region is due to the presence $f_0(980)$ resonance in the amplitude $S_1$. 

The non-flip amplitudes $|S_0(ij)|^2$ are very small and nearly constant. As a result we find $|S_1(ij)|^2 \approx I_S(i,j)$ for all $i,j=1,2$. Using (4.9) this means that
\begin{equation}
|S_1^0|^2 \approx |S_1|^2
\end{equation}
There is also a clear similarity between the phase $\Phi(L_1)-\Phi(S_1)$ and the phases $\Phi_{LS}$ and $\overline{\Phi}_{LS}$ which implies that 
\begin{equation}
\Phi(L_1^0)-\Phi(S_1^0) \approx \Phi(L_1)-\Phi(S_1)
\end{equation}
In equations (6.28) and (6.29) the Solutions $i=1,2$ on the l.h.s. correspond to Solutions $ij=11,22$ on the r.h.s., respectively. The extrapolation of $S_1$ to $\rho^0_y=0$ obtained in the analysis of unpolarazied target data carries nearly as much information as the exact amplitude from analysis of polarized target data.

\section{$\rho^0(770)-f_0(980)$ mixing and Lorentz symmetry.}

It is generally expected that the position and the width of the $\rho^0(770)$ peak as observed in the spin averaged cross-section will be faithfully reproduced on the level of spin amplitudes. The CERN data on polarized target show that this is not the case. From Figures 2, 11 and 12 we see that the $\rho^0(770)$ production is suppressed in all target spin "up" amplitudes while the target spin "down" spectra dominate the $\rho^0(770)$ production. The width at half-height of the peaks in the longitudinal spectra  $|L_u|^2$ and $|L_d|^2$ is the expected $\sim 150$ MeV. However, the $\rho^0(770)$ width shows different values in different transverse spectra. In the spin "down" spectra the $\rho^0(770)$ width is narrower at $\sim 120$ MeV in $|U_d|^2$ but wider at $\sim 180$ MeV in $|N_d|^2$. These values are reversed in the spin "up" spectra with $\sim 180$ MeV and $\sim 120$ MeV in amplitudes $|U_u|^2$ and $|N_u|^2$, respectively. Such large variations in the $\rho^0(770)$ width are entirely unexpected and appear anomalous.

\begin{figure}
\includegraphics[width=12cm,height=10.5cm]{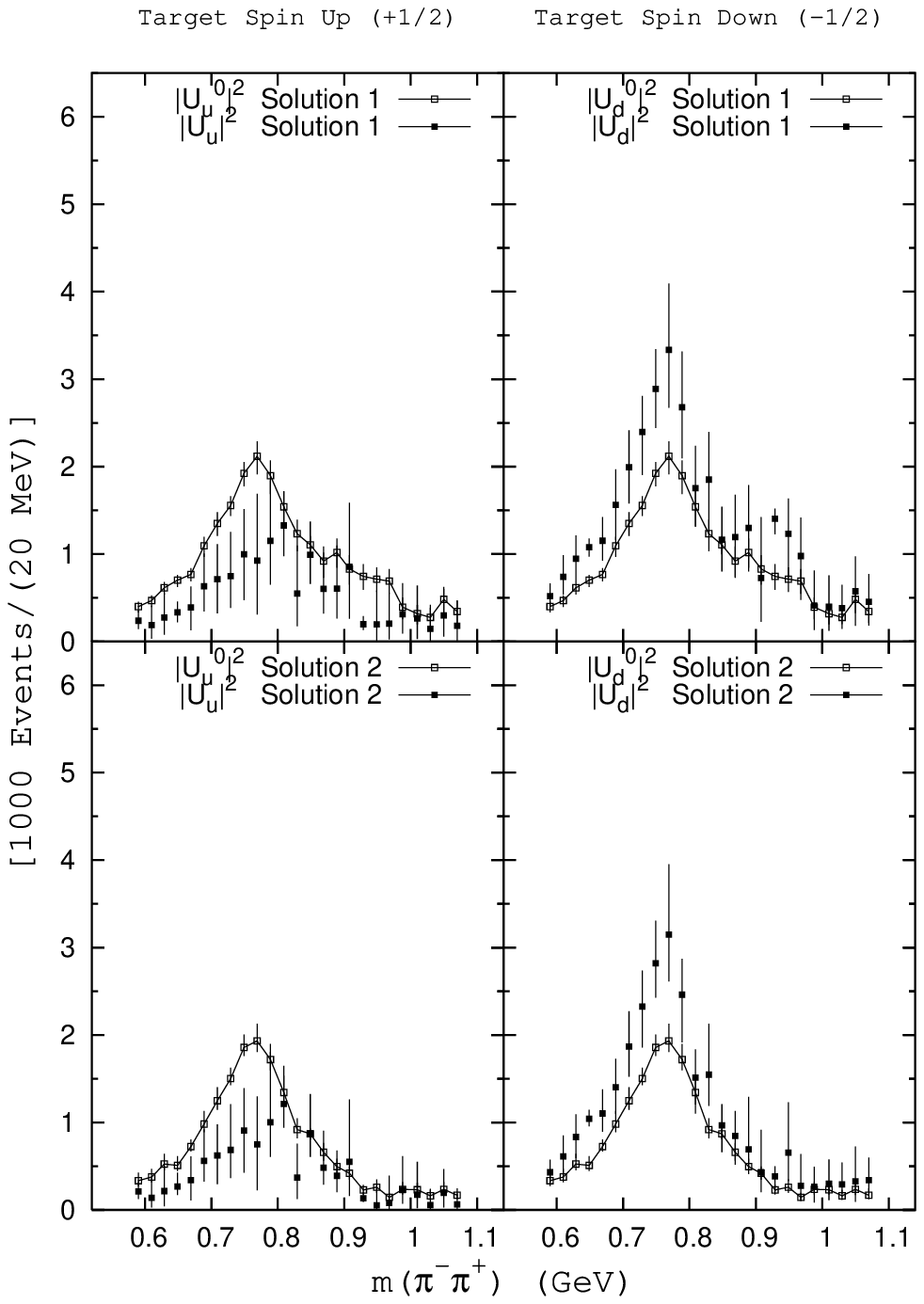}
\caption{Moduli of $P$-wave amplitudes $|U_\tau|^2$ and $|U_\tau^0|^2$ from Analyses I and II (with line).}
\label{Figure 11.}
\end{figure}

\begin{figure}
\includegraphics[width=12cm,height=10.5cm]{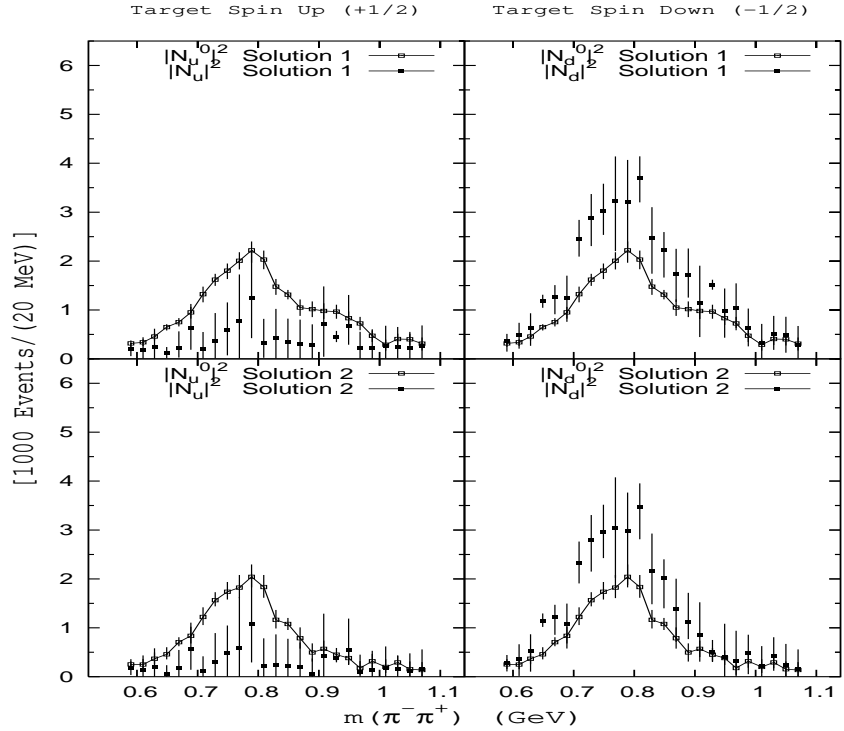}
\caption{Moduli of $P$-wave amplitudes $|N_\tau|^2$ and $|N_\tau^0|^2$ from Analyses I and II (with line).}
\label{Figure 12.}
\end{figure}

The rotational symmetry of strong interactions requires that the width and mass of a resonance do not depend on its helicity. It also prevents the mixing of scalar and vector resonances in the same partial wave with a definite spin $J$. If the rotational symmetry was broken in pion production then the widths of $\rho^0(770)$ in the $P$-wave amplitudes $L_\tau$, $U_\tau$ and $N_\tau$ would differ and $\rho^0(770)-f_0(980)$ mixing could occur in both $S$- and $P$-wave amplitudes.

To test the rotational symmetry we need information on transversity amplitudes $H_\tau^\lambda$ with definite dipion helicity $\lambda=0,\pm1$. The transverse amplitudes $U_\tau$ and $N_\tau$ are a mix of transverse amplitudes with dipion helicities $\lambda$=+1 and -1, and thus are not suitable to test the rotational symmetry. The required amplitudes with transverse helicities $H_\tau^{+1}$ and $H_\tau^{-1}$ are related to the amplitudes $U_\tau$ and $N_\tau$~\cite{lutz78,svec12b}
\begin{eqnarray}
H^{+1}_\tau & = & {1 \over {\sqrt{2}}}(U_\tau + N_\tau)\\
\nonumber
H^{-1}_\tau & = & {1 \over {\sqrt{2}}}(U_u - N_u)
\end{eqnarray}
Their partial wave intensities and polarizations can be calculated from the data on polarized target
\begin{eqnarray}
I(H_\tau) & = & |H_\tau^{+1}|^2 + |H_\tau^{-1}|^2 = |U_\tau|^2 + |N_\tau|^2\\
\nonumber
P(H_\tau) & = & 2Re (H_\tau^{+1}H_\tau^{-1*} ) = |U_\tau|^2 - |N_\tau|^2
\end{eqnarray}
For amplitudes with zero helicity we have longitudinal amplitudes $L_\tau$. It is convenient to relabel them as $H^0_\tau \equiv L_\tau$. Their moduli squared shown in Figure 2 are then the longitudinal intensities 
\begin{equation}
I(H^0_\tau) = |H^0_\tau|^2
\end{equation}

\begin{figure}
\includegraphics[width=12cm,height=10.5cm]{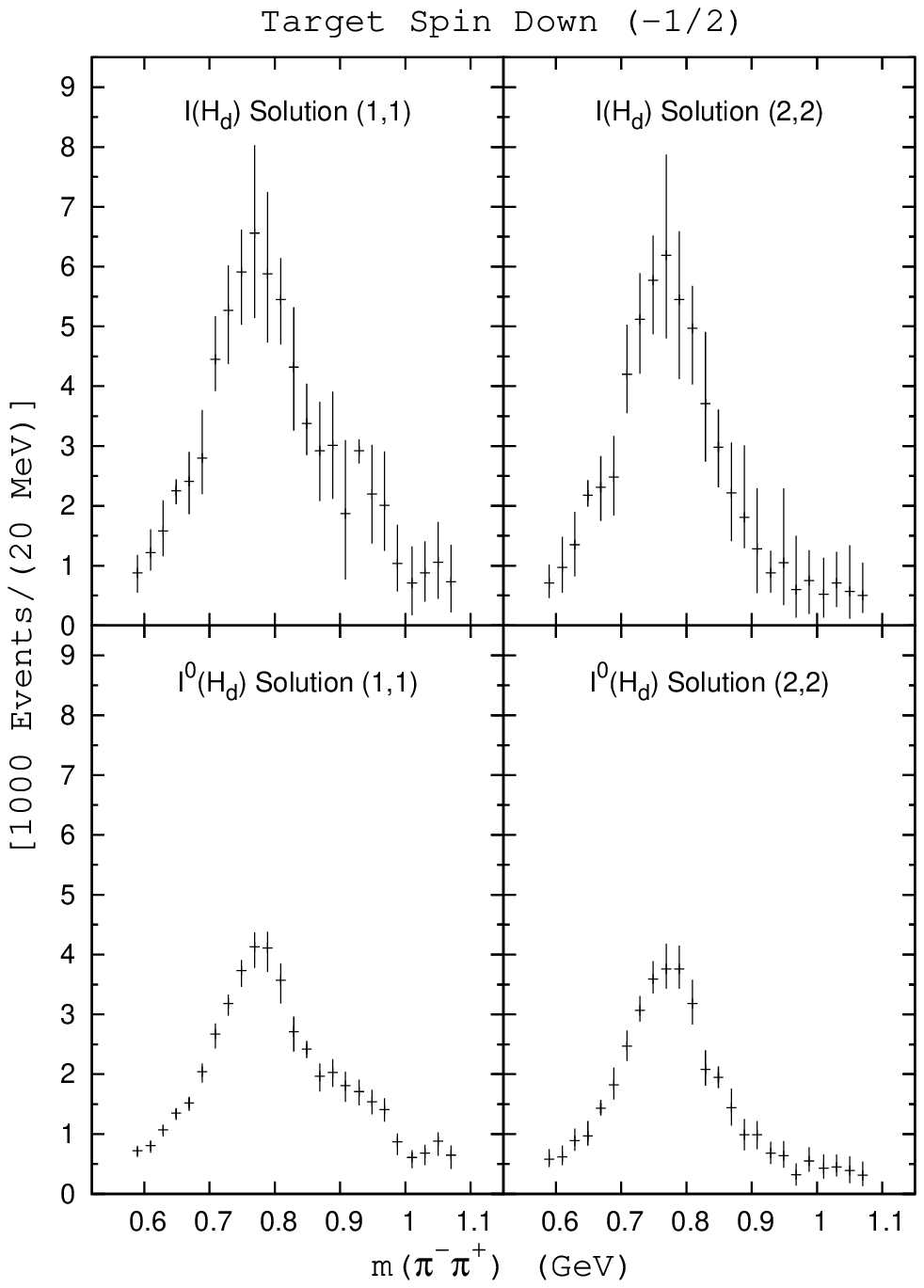}
\caption{Solutions (1,1) and (2,2) for the intensities $I(H_d)$ and $I^0(H_d)$ from Analyses I and II.}
\label{Figure 13.}
\end{figure}

\begin{figure}
\includegraphics[width=12cm,height=10.5cm]{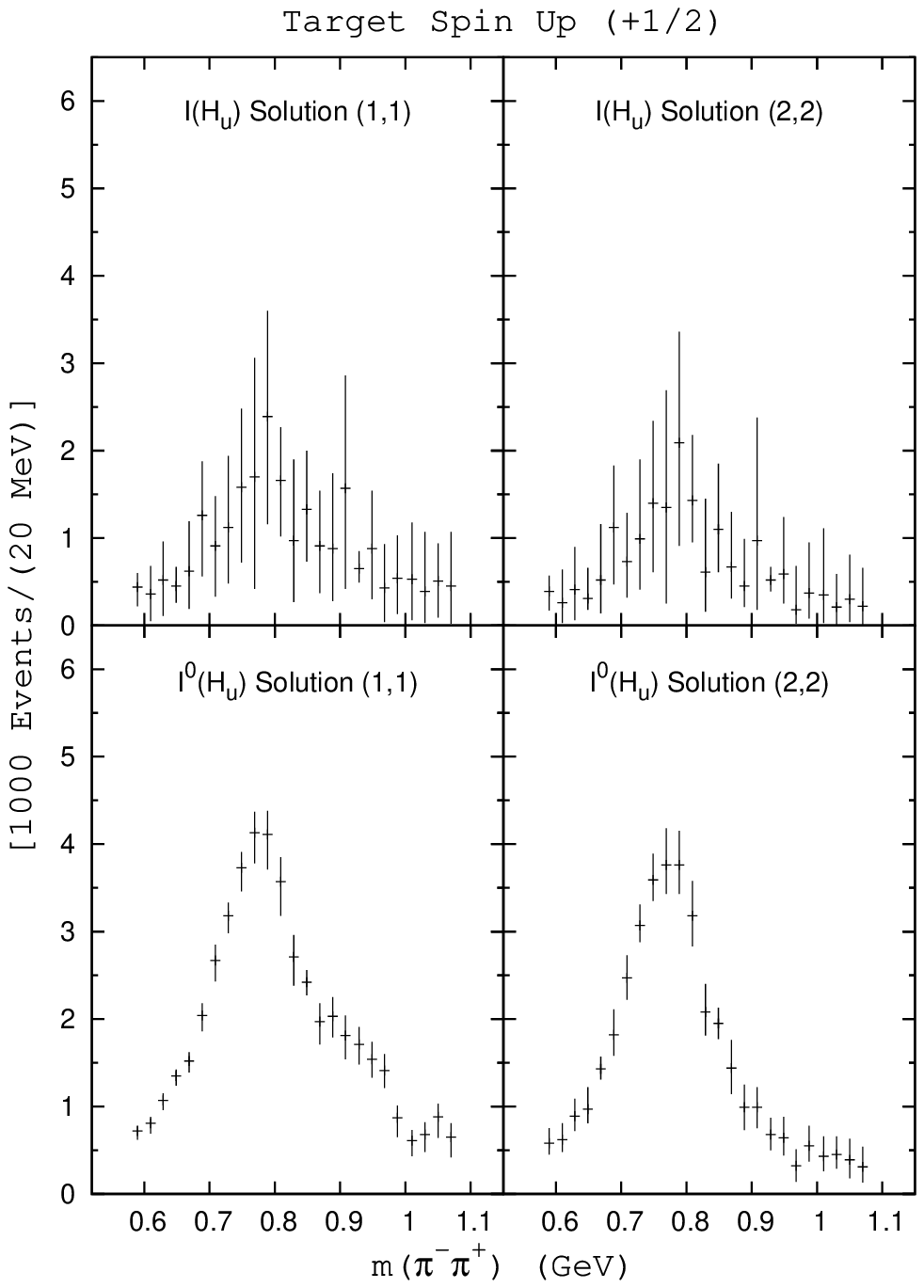}
\caption{Solutions (1,1) and (2,2) for the intensities $I(H_u)$ and $I^0(H_u)$ from Analyses I and II.}
\label{Figure 14.}
\end{figure}

\begin{figure}
\includegraphics[width=12cm,height=10.5cm]{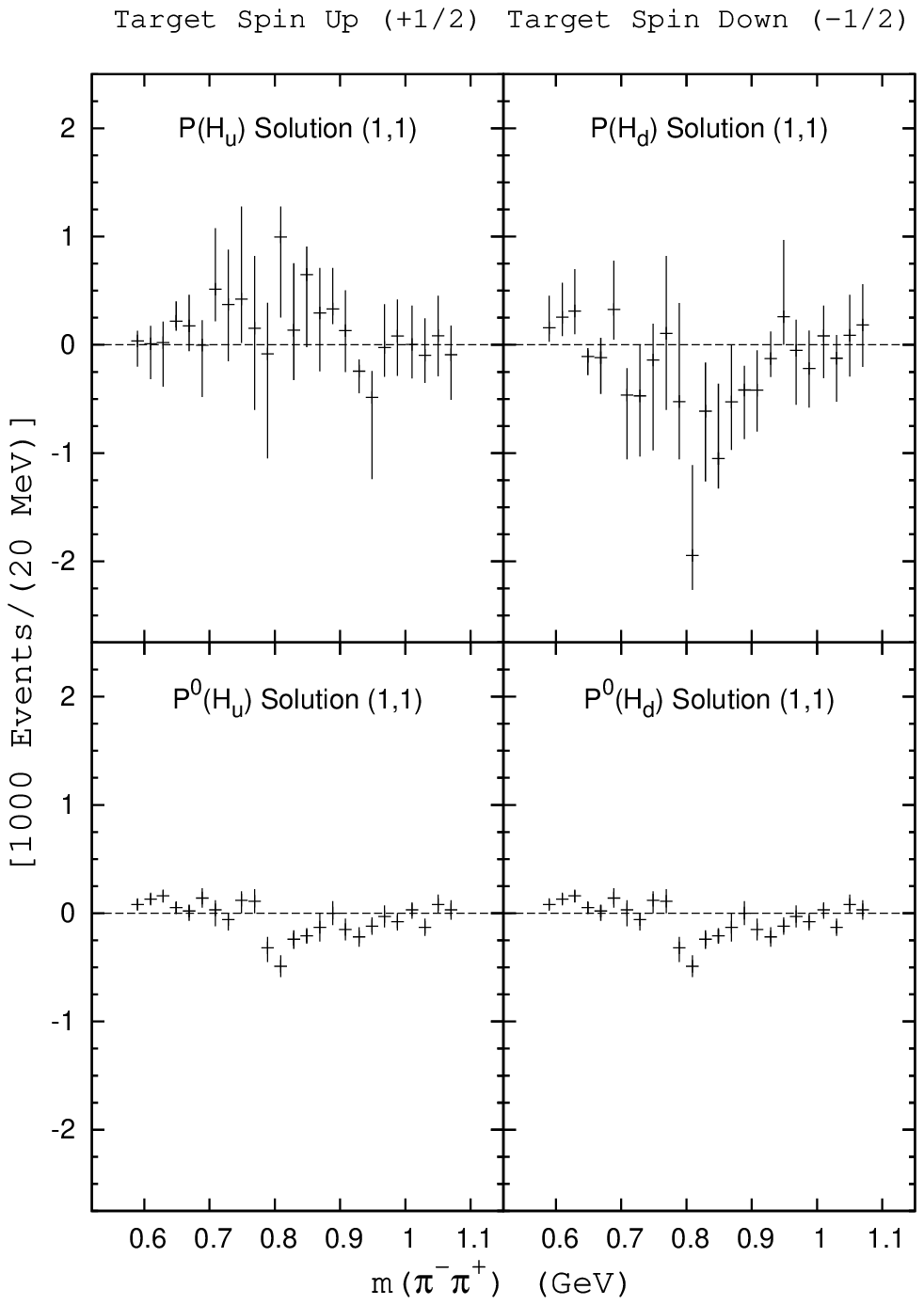}
\caption{Polarizations $P(H_\tau)$ and $P^0(H_\tau)$ from Analyses I and II. Solutions (1,1) and (2,2) are equal.}
\label{Figure 15.}
\end{figure}

Figures 13 and 14 show the transverse intensities $I(H_d)$ and  $I(H_u)$ for two solutions (1,1) and (2,2). The intensities show a clear peak with the same width at half-height of $\sim$ 150 MeV for both target spins in both solutions. This transverse $\rho^0(770)$ width is exactly the same as the longitudinal width at half-height in the intensities $I(H^0_d)$ and $I(H^0_u)$. This indicates that the $\rho^0(770)$ poles in all helicity amplitudes $H^{\lambda}_u, H^{\lambda}_d$, $\lambda=0, \pm1$ have the same width $\sim$ 150 MeV.

To explain the large differences in the $\rho^0(770)$ witdth observed in the spectra $|U_d|^2$ and $|N_d|^2$ we note from (7.2)
\begin{eqnarray}
|U_\tau|^2 & = & {1 \over {2}}( I(H_\tau) + P(H_\tau) )\\
\nonumber
|N_\tau|^2 & = & {1 \over {2}}( I(H_\tau) - P(H_\tau) )
\end{eqnarray}
Figure 15 shows the polarizations $P(H_\tau), \tau=u,d$ which are equal for both Solutions (1,1) and (2,2). The polarization $P(H_d)$ is negative and broad in the $\rho^0(770)$ mass region. As the result of the opposite signs of $P(H_d)$ in (7.4), the apparent $\rho^0(770)$ width is narrower in $|U_d|^2$ and wider in $|N_d|^2$. The polarization $P(H_u)$ is positive and broad around the $\rho^0(770)$ mass and therefore it has an opposite effect on their apparent widths, as observed. Thus the difference in the resonance apparent widths in the amplitudes $|U_\tau|^2$ and $|N_\tau|^2$ is entirely due to the interference of the amplitudes $H^{+1}_d$ and $H^{-1}_d$.

Figures 11 and 12 show also the results for amplitudes $|U_\tau^0|^2$ and 
$|N_\tau^0|^2$ from the Analysis II. It is remarkable that in this analysis the resonance widths at half-height in both amplitudes and in both Solutions 1,2 are the same at $\sim 150$ MeV. From (7.4) this implies $P^0(H_\tau)=|U^0_\tau|^2-|N^0_\tau|^2 \approx 0$, as observed in Figure 15. The equal widths are therefore observed also in both solutions (1,1) and (2,2) for the transverse intensities $I^0(H_d)=|U_d^0|^2+|N_d^0|^2=I^0(H_u)$ in Figures 13 and 14.

\section{Interpretation of the observed $\rho^0(770)-f_0(980)$ spin mixing.}

Vector-scalar meson mixing is not a new idea. In 1977 Chin studied $\sigma(500)-\omega(783)$ mixing in nucleon-nucleon interactions in high density matter~\cite{chin77}. In 2000-2002 Gale and collaborators examined the effects of $\rho(770)-a_0(980)$ mixing on the dilepton production in relativistic heavy ion collisions~\cite{gale00,gale01,gale02a,gale02b} measured at RHIC. The $\sigma(500)-\omega(783)$ mixing was induced by the ground state of the system of the interacting nucleons, while the $\rho(770)-a_0(980)$ mixing originated in nucleon-nucleon excitations in the medium of nuclear matter. In both cases there was no violation of fundamental symmetries because the interaction Lagrangian conserved those symmetries including the Lorentz symmetry. The meson spin mixing was due to the interaction of nucleon-nucleon scattering process with its environment.

We conclude from the tests in the Section VII. of the rotational/Lorentz symmetry in the Analyses I and II that the $\rho^0(770)-f_0(980)$ spin mixing is fully consistent with the conservation of rotational/Lorentz symmetry by the production mechanism. This remarkable consistency suggests that the mechanism for $\rho^0(779)-f_0(980)$ spin mixing and the production mechanism have different dynamical origins. While there is no reason to assume that the production mechanism is governed by anything other than the $S$-matrix dynamics, the $\rho^0(770)-f_0(980)$ mixing must originate in a new spin mixing interaction outside of the Standard Model since no fundamental interaction of the Standard Model mixes particle spins. The consistency suggests that the produced $S$-matrix final state $\rho_f(S)$ interacts with a quantum environment $E$ to produce the observed spin mixing final state $\rho_f(O)$. The observed amplitudes are not the same as the $S$-matrix amplitudes. This new non-standard interaction spontaneously violates rotational/Lorentz symmetry in the observed amplitudes while conserving this symmetry in the $S$-matrix amplitudes and in the Standard Model Lagrangian. 

Recall that the production and decay of a resonance of spin $J_R$ and isospin $I_R$ are both fully described by the $S$-matrix partial wave production amplitudes $U^{J_R}_{\lambda \tau}$ and $N^{J_R}_{\mu \tau}$. The spin and the isospin of the resonance are conserved within these $S$-matrix amplitudes. Unitary $S$-matrix amplitudes do not mix resonances of different spins in the same partial wave since none of the fundamental interactions in the Standard Model mixes states of different spins. The appearance of a resonance in a mixed partial wave production amplitude with $J \neq J_R$, $I \neq I_R$ is not a violation of the symmetries and conservation laws of the Standard Model because it originates in a new dynamics from the outside of the Standard Model within the  observed mixed amplitudes. The mixing of $S$-matrix partial wave production amplitudes with different spins to form new observable partial wave production amplitudes is a new phenomenon outside of Standard Model and represents a genuine new physics.

\section{Conclusions and outlook.}

We have presented experimental evidence for $\rho^0(770)-f_0(980)$ mixing in $S$- and $P$-wave transversity amplitudes in $\pi^- p \to \pi^- \pi^+ n$ from both polarized and unpolarized target data measured at CERN. A model independent determination of $S$-and $P$-wave helicity amplitudes confirms the evidence from transversity amplitudes and specifies the relative phase $\omega$ of amplitudes with opposite transversities. All three analyses yield mutually consistent results. We show that the measurements of $Im \rho^0_x$ and $Im \rho^0_z$ resolve the sign ambiguity of relative phases. As a result, there are four sets of transversity amplitudes and four sets of helicity amplitudes. The presented results are in agreement with all other amplitude analyses of the five measurements of pion production on polarized targets. A full review of the evidence for $\rho^0(770)-f_0(980)$ mixing from these other analyses is given in Ref.~\cite{svec12d}.

The principal difference between the Analyses I and II is the spin dependence of the moduli $|A_\tau|^2$. In the Analysis I the spin $up$ moduli $|A_u|^2$ are suppressed compared to the spin $down$ moduli $|A_d|^2$ while in the Analysis II these moduli are equal $|A_u|^2=|A_d|^2$. The suppression of $|A_u|^2$ in the Analysis I is a part of a more complex spin dependence of the amplitudes $|A_\tau|^2$ which show oscillations as a function of momentum transfer $t$ in the $\rho^0(770)$ mass region~\cite{svec90}. 

The amplitude Analyses I and II provide the first direct experimental evidence that the width of a resonance does not depend on its helicity as expected from the rotational symmetry of strong interactions. Since the observed $\rho^0(770) - f_0(980)$ mixing is consistent with rotational/Lorentz symmetry of the production mechanism and since the fundamental interactions of the Standard Model do not mix particles with different spins, the $\rho^0(770) - f_0(980)$ mixing must arise from a new kind of interaction independently involved in the pion creation process. The consistency itself suggests that the spin mixing mechanism arises from an interaction of the produced final state $\rho_f(S)$ with a quantum environment. In two related papers~\cite{svec12b,svec13b} we present evidence for the existence of the quantum environment and its pure dephasing interaction with the produced $S$-matrix final state $\rho_f(S)$. 

Dark matter and dark energy are two omnipresent environments in the Universe with a non-standard interaction with baryonic matter. Consistency of the quantum environment with the Standard Model implies that it is a universal environment in the Universe with a non-standard interaction with baryonic matter. This similarity suggests that the quantum environment could be identified with the dark matter, or as the common origin of dark matter and dark energy. Dedicated measurements of meson production processes on polarized targets are necessary to confirm and explore the spin mixing phenomena and to clarify their possible connection to cosmology.

\appendix

\section*{Determination of the phases of natural exchange amplitudes and the resolution of sign ambiguities in the phases $\Phi_{SL}$ and $\overline{\Phi}_{SL}$.} 

The $S$- and $P$-wave density matrix elements $Im \rho^0_x$ and $Im \rho^0_z$ involve interferences between reduced unnatural and natural exchange amplitudes of opposite transversity shown in Table I.. First we shall show that the elements 
$Im (\rho^0_x)^{01}_{s1}$ and $Im (\rho^0_z)^{01}_{s1}$ determine the phases $\alpha_N=\Phi_{N_u}-\Phi_{S_d}$ and $\overline{\alpha}_N=\Phi_{N_d}-\Phi_{S_u}$ of the natural exchange amplitudes $N$ and $\overline {N}$, respectively, with a two-fold ambiguity for each set of moduli $|N(i)|,|\overline {N}(j)|,i,j,=1,2$. From the Table I. and (2.6) we have
\begin{equation}
r_1=\sqrt{2}Im (\rho^0_x)^{01}_{s1} \Sigma = Re(-|S|\overline {N}^*+N|\overline{S}|)=
-|S||\overline {N}| \cos \overline {\alpha}_N+|N||\overline {S}| \cos \alpha_N
\end{equation}
\[ 
s_1=\sqrt{2}Im (\rho^0_z)^{01}_{s1} \Sigma = Im(+|S|\overline {N}^*-N|\overline{S}|)=
-|S||\overline {N}| \sin \overline {\alpha}_N-|N||\overline {S}| \sin \alpha_N
\]
where $\Sigma = d^2 \sigma /dmdt$. From (1) we can solve for $\cos \alpha_N$ and $\sin \alpha_N$
\begin{eqnarray}
\cos \alpha_N & = & {1 \over{|N||\overline{S}|}} \Bigl (r_1+|S||\overline {N}| \cos \overline {\alpha}_N \Bigr ) \\
\nonumber 
\sin \alpha_N & = & {-1 \over{|N||\overline{S}|}} \Bigl (s_1+|S||\overline {N}| \sin \overline {\alpha}_N \Bigr ) 
\nonumber 
\end{eqnarray}
Substituting into $\cos^2 \alpha_N+\sin^2 \alpha_N=1$ we find
\begin{equation}
\cos \overline {\alpha}_N={1\over{2|S||\overline {N}|r_1}} \Bigl ( A-2|S||\overline{N}|s_1 \sin \overline {\alpha}_N \Bigr )
\end{equation}
where
\[
A=|N|^2|\overline {S}|^2 -|S|^2|\overline {N}|^2-(r_1^2+s_1^2)
\]
Substituting into $\cos^2 \overline {\alpha}_N+\sin^2 \overline {\alpha}_N=1$ yields a quadratic equation for $\sin \overline {\alpha}_N$ with two solutions
\begin{equation}
\sin \overline {\alpha}_N = {A \over {2|S||\overline {N}|(r_1^2+s_1^2)}} 
\Bigl ( s_1 \pm r_1 \sqrt{B-1} \Bigr )
\end{equation}
where
\[
B={{4|S|^2| \overline{N}|^2(r_1^2+s_1^2)} \over {A^2}}
\]
From (4) we can now calculate $\cos \overline {\alpha}_N$
\begin{equation}
\cos \overline {\alpha}_N= {A \over {2|S||\overline {N}|(r_1^2+s_1^2)}} 
\Bigl ( r_1 \mp s_1 \sqrt{B-1} \Bigr ) 
\end{equation}
and from (3) we obtain
\begin{eqnarray}
\cos \alpha_N & = &  {A \over {2|N||\overline {S}|(r_1^2+s_1^2)}} 
\Bigl ( r_1C \mp s_1 \sqrt{B-1} \Bigr )\\
\nonumber
\sin \alpha_N & = &  {{-A} \over {2|N||\overline {S}|(r_1^2+s_1^2)}} 
\Bigl ( s_1C \pm r_1 \sqrt{B-1} \Bigr )
\nonumber
\end{eqnarray}
where
\[
C={{A+2(r_1^2+s_1^2)} \over {A}}
\]
Note that for each set of moduli $|A(i)|,|\overline{A}(j)|$, $i,j=1,2$ the two solutions for the phases $\alpha_N(ij,\sigma), \overline {\alpha}_N(ij,\sigma)$ depend $i,j$ and the sign $\sigma=\pm 1$ in (4) but do not depend the sign ambiguities $\epsilon, \overline{\epsilon}$ of amplitudes $A=L,U$ in (3.19). As a result, the amplitudes $N(ij,\sigma)$ and $\overline{N}(ij,\sigma)$ do not depend on signs $\epsilon, \overline{\epsilon}$.

We shall now show that for each set of the moduli $i,j$ only one solution $N(ij,\sigma)$ and $\overline{N}(ij,\sigma)$ is consistent with the remaining equations from the Table I. and that this solution resolves the four-fold sign ambiguity in relative phases of unnatural exchange amplitudes $L$ and $U$. To this end we introduce a convenient notation for real and imaginary parts of reduced transversity amplitudes $A=L,U,N$ 
\begin{eqnarray}
A=A_1+iA_2 & = & |A| \cos \alpha_A + i|A| \sin \alpha_A\\
\nonumber
\overline {A} =\overline {A}_1+i\overline {A}_2 & = & 
|\overline {A}| \cos \overline {\alpha}_A+i|\overline {A}| \sin \overline {\alpha}_A
\nonumber
\end{eqnarray}
where the phases $\alpha_A=\alpha_A(i,\epsilon)$ and $\overline {\alpha}_A=\overline {\alpha}_A(j,\overline{\epsilon})$ for amplitudes $A=L,U$ are given by (3.19). The equations for the remaining density matrix elements from the Table I. then take the form
\begin{eqnarray}
r_2= \sqrt{2} Im ( \rho^0_x)^{11}_{01} \Sigma & = & 
\overline {L}_1 N_1+ \overline {L}_2 N_2-L_1 \overline {N}_1 -L_2 \overline {N}_2 \\
\nonumber
s_2= \sqrt{2} Im ( \rho^0_z)^{11}_{01} \Sigma & = & 
\overline {L}_2 N_1- \overline {L}_1 N_2+L_2 \overline {N}_1 -L_1 \overline {N}_2 
\nonumber
\end{eqnarray}
\begin{eqnarray}
r_3=-Im ( \rho^0_x)^{11}_{-11} \Sigma & = & 
\overline {U}_1 N_1+ \overline {U}_2 N_2-U_1 \overline {N}_1 -U_2 \overline {N}_2 \\
\nonumber
s_3=-Im ( \rho^0_z)^{11}_{-11} \Sigma & = & 
\overline {U}_2 N_1- \overline {U}_1 N_2+U_2 \overline {N}_1 -U_1 \overline {N}_2 
\nonumber
\end{eqnarray}

The equations (8) and (9) are four linear equations for four unknowns $N_1,N_2,\overline{N}_1, \overline{N}_2$ with four solutions $N_1(ij,\epsilon,\overline{\epsilon})$, $N_2(ij,\epsilon,\overline{\epsilon})$, $\overline{N}_1(ij,\epsilon,\overline{\epsilon})$ and $\overline{N}_2(ij,\epsilon,\overline{\epsilon})$ corresponding to $\epsilon=\pm1$ and $\overline{\epsilon}=\pm1$. One of these four solutions of (8) and (9) must be equal to one of the solutions of (1). To be specific, let us suppose that this physical solution for $N(ij,\sigma), \overline {N}(ij,\sigma)$ is the solution with the positive sign $\sigma=+1$ in (4). It corresponds to some specific signs $\epsilon^+$ and $\overline{\epsilon}^+$ so let us label all amplitudes in this solution $A^+,\overline {A}^+$. We now show that the amplitudes $N^-,\overline {N}^-$ for the other solution with $\sigma=-1$ cannot be a solution of (8) and (9) for any choice of signs $\epsilon$ and $\overline{\epsilon}$ of the phases in (3.19).

To prove this statetement let us suppose the contrary and assume that the amplitudes $N^-,\overline {N}^-$ satisfy (8) and (9) for some amplitudes $A^-,\overline{A}^-$, $A=L,U$ from (3.19) corresponding to particular values of signs $\epsilon^-$ and $\overline{\epsilon}^-$. Since the amplitudes $A^-,\overline{A}^-$, $A=L,U$ differ from the amplitudes $A^+,\overline{A}^+$, $A=L,U$ only in the sign of phases, their real parts are the same and the imaginary parts differ at most by sign
\begin{equation}
A_2^-=\lambda A_2^+, \quad \overline {A}_2^-=\overline{ \lambda} \overline {A}_2^+, \quad A=L,U
\end{equation}
where $\lambda= \pm 1, \overline {\lambda}= \pm 1$. In the next step we subtract the set of equations with amplitudes $N^-,\overline {N}^-$ from the set with amplitudes $N^+,\overline {N}^+$ and get a homogeneous set of equations
\begin{eqnarray}
\overline {L}_1^+(N_1^+-N_1^-)+\overline{L}_2^+(N_2^+-\overline{\lambda}N_2^-)-
L_1^+(\overline {N}_1^+-\overline {N}_1^-)-L_2^+(\overline{N}_2^+-\lambda \overline {N}_2^-) & = & 0\\
\nonumber
\overline {L}_2^+(N_1^+-\overline{\lambda}N_1^-)-\overline{L}_1^+(N_2^+-N_2^-)+
L_2^+(\overline {N}_1^+-\lambda \overline {N}_1^-)-L_1^+(\overline{N}_2^+-\overline {N}_2^-) & = & 0
\nonumber
\end{eqnarray}
\begin{eqnarray}
\overline {U}_1^+(N_1^+-N_1^-)+\overline{U}_2^+(N_2^+-\overline{\lambda}N_2^-)-
U_1^+(\overline {N}_1^+-\overline {N}_1^-)-U_2^+(\overline{N}_2^+-\lambda \overline {N}_2^-) & = & 0\\
\nonumber
\overline {U}_2^+(N_1^+-\overline{\lambda}N_1^-)-\overline{U}_1^+(N_2^+-N_2^-)+
U_2^+(\overline {N}_1^+-\lambda \overline {N}_1^-)-U_1^+(\overline{N}_2^+-\overline {N}_2^-) & = & 0
\nonumber
\end{eqnarray}
We now write (6) in the form
\begin{eqnarray}
\cos \alpha^\pm_N & = & {1\over {|N|}} \bigl ( a_1r_1 \mp a_2s_1 \bigr )\\
\nonumber
\sin \alpha^\pm_N & = & {1\over {|N|}} \bigl ( a_1s_1 \pm a_2r_1 \bigr )
\nonumber
\end{eqnarray}
and calculate differences and sums
\begin{equation}
N_1^+-N_1^- = -2a_2s_1, \quad N_1^++N_1^- = +2a_1r_1
\end{equation}
\[
N_2^+-N_2^- = +2a_2r_1, \quad N_1^++N_1^- = +2a_1s_1
\]
We can write similar expressions for amplitudes $\overline{N}^+$ and $\overline {N}^-$ by replacing the paramerters $a_1,a_2$ in (14) with parameters $\overline {a}_1, \overline {a}_2$ corresponding to the form (13) of the equations (5) and (4) for
 $\cos \overline {\alpha}_N$ and $\sin \overline {\alpha}_N$.

We now examine the equations (11) and (12) for each possible choice of $\lambda$ and $\overline {\lambda}$. The case $\lambda=\overline {\lambda}=+1$ is excluded as the 
amplitudes $N^-,\overline {N}^-$ cannot satisfy the same system as the amplitudes  
$N^+,\overline {N}^+$. For the case with $\lambda=\overline {\lambda}=-1$ the equations (11) and (12) take the form
\begin{eqnarray}
s_1 \Bigl (-\overline {L}_1^+a_2+\overline {L}_2^+a_1 +
L_1^+ \overline{a}_2 -L_2^+\overline {a}_1 \Bigr ) & = & 0\\
\nonumber
r_1 \Bigl (-\overline {L}_1^+a_2+\overline {L}_2^+a_1 -
L_1^+ \overline{a}_2 +L_2^+\overline {a}_1 \Bigr ) & = & 0\\
s_1 \Bigl (-\overline {U}_1^+a_2+\overline {U}_2^+a_1 +
U_1^+ \overline{a}_2 -U_2^+\overline {a}_1 \Bigr ) & = & 0\\
\nonumber
r_1 \Bigl (-\overline {U}_1^+a_2+\overline {U}_2^+a_1 -
U_1^+ \overline{a}_2 +U_2^+\overline {a}_1 \Bigr ) & = & 0
\nonumber
\end{eqnarray}
The terms in the parentheses are all different. In particular, the large differences between the moduli $|L|,|\overline {L}|$ and $|U|,|\overline {U}|$ mean large differences in the parentheses for $r_1$ at least one of which must be non-zero. Since $r_1$ has been measured in the CERN experiments on polarized targets and it is non-zero, the equations (15) and (16) cannot be satisfied and this case is excluded.

For the case $\lambda=+1$ and $\overline {\lambda}=-1$ the equations  (11) and (12) read
\begin{eqnarray}
\Bigl (-\overline {L}_1^+a_2+\overline {L}_2^+a_1 +
L_1^+ \overline{a}_2 \Bigr )s_1 -\Bigl (L_2^+\overline {a}_2 \Bigr )r_1 & = & 0\\
\nonumber
\Bigl (-\overline {L}_1^+a_2+\overline {L}_2^+a_1 -
L_1^+ \overline{a}_2 \Bigr )r_1 -\Bigl (L_2^+\overline {a}_2 \Bigr )s_1 & = & 0\\
\Bigl (-\overline {U}_1^+a_2+\overline {U}_2^+a_1 +
U_1^+ \overline{a}_2 \Bigr )s_1 -\Bigl (U_2^+\overline {a}_2 \Bigr )r_1 & = & 0\\
\nonumber
\Bigl (-\overline {U}_1^+a_2+\overline {U}_2^+a_1 -
U_1^+ \overline{a}_2 \Bigr )r_1 -\Bigl (U_2^+\overline {a}_2 \Bigr )s_1 & = & 0
\nonumber
\end{eqnarray}
Combining the first two and the last two equations we obtain two equations of interest
\begin{eqnarray}
\Bigl (-\overline {L}_1^+a_2+\overline {L}_2^+a_1 +L_1^+ \overline{a}_2 \Bigr )s_1^2 & = & 
\Bigl (-\overline {L}_1^+a_2+\overline {L}_2^+a_1 -L_1^+ \overline{a}_2 \Bigr )r_1^2\\
\Bigl (-\overline {U}_1^+a_2+\overline {U}_2^+a_1 +U_1^+ \overline{a}_2 \Bigr )s_1^2 & = & 
\Bigl (-\overline {U}_1^+a_2+\overline {U}_2^+a_1 -U_1^+ \overline{a}_2 \Bigr )r_1^2
\nonumber
\end{eqnarray}
The terms in the parentheses are all different. The parentheses for amplitudes $L, \overline {L}$ and amplitudes $U, \overline {U}$ are not proportional to each other since the phases of $L, \overline {L}$ and $U, \overline {U}$ are 180$^o$ out of phase, as seen in Fig. 7. Moreover, as we have discussed in Section VII., the moduli of these amplitudes have different $\rho^0(770)$ widths and structures around $f_0(980)$. The equations (17) and (18) thus cannot be satisfied and this case is excluded. The analysis of the case $\lambda=-1$ and $\overline{\lambda}=+1$ is similar with the same conclusion. The solution $N^+, \overline {N}^+$ selects a unique set of phases of amplitudes $L,U$ and $\overline{L}, \overline {U}$ corresponding to $\epsilon^+$ and $\overline{\epsilon}^+$. The equations (8) and (9) change when the phases of these amplitudes change sign. The change results in a different solution for amplitudes $N, \overline {N}$ which is not compatible with the data in (1).

We conclude that the measurements of $S$- and $P$-wave density matrix elements $Im \rho^0_x$ and $Im \rho^0_z$ unambiguously select a unique solution for $S$- and $P$-wave reduced transversity amplitudes. For any set $i,j$ of solutions for the moduli there exists only one solution for the phases of natural exchange amplitudes and only one set of signs of phases of unnatural exchange amplitudes in (3.19). The corresponding solution of equations (8) and (9) for $N_1,N_2,\overline{N}_1, \overline{N}_2$ is the only solution that satisfies the conditions $|N(i)|^2=N_1^2+N_2^2$ and $|N(j)|^2=\overline{N}_!^2+\overline{N}_2^2$.



\begin{thebibliography} {}

\bibitem{erwin61} A.R.~Erwin, R.~March, W.D.~Walker and E.~West, {\sl Evidence for a $\pi \pi$ Resonance in the $I=1$, $J=1$ State}, Phys.Rev.Lett. {\bf 6}, 628 (1961).

\bibitem{hagopian63} V.~Hagopian and W.~Selow, {\sl Experimental Evidence on $\pi \pi$ Scattering Near the $\rho^0$ and $f^0$ Resonances from $\pi^- + p \to \pi + \pi + nucleon$   at 3 BeV/c}, Phys.Rev.Lett. {\bf 10}, 533 (1963).

\bibitem{islam64} M.M.~Islam and R.~Pinn, {\sl Study of $\pi^- \pi^+$ System in $\pi^- + p \to \pi^- + \pi^+ + n$ Reaction}, Phys.Rev.Lett. {\bf 12}, 310 (1964).

\bibitem{patil64} S.H.~Patil, {\sl Analysis of the $S$-wave in $\pi \pi$ Interactions}, Phys.Rev.Lett. {\bf 13}, 261 (1964).

\bibitem{durand65} L.~Durand III and Y.T.~Chiu, {\sl Decay of the $\rho^0$ Meson and the Possible Existence of a $T=0$ Scalar Di-Pion}, Phys.Rev.Lett. {\bf 14}, 329 (1965).

\bibitem{baton65} J.P.~Baton {\sl et al.}, {\sl Single Pion Production in $\pi^-p$ Interactions at 2.75 GeV/c}, Nuovo Cimento {\bf 35}, 713 (1965).

\bibitem{apel72} W.D.~Apel {\sl et al.}, {\sl Results on $\pi \pi$ Interaction in the Reaction $\pi^- p \to \pi^0 \pi^0 n$ at  8 GeV/c}, Phys.Lett. {\bf B41}, 542 (1972).

\bibitem{pennington73} M.R.~Pennington and S.D.~Protopopescu, {\sl How Roy's Equations Resolve Up-Down Ambiguity and Reproduce $S^*$ Resonance}, Phys.Rev. {\bf D7},2591 (1973).

\bibitem{lutz78} G.~Lutz and K.~Rybicki, {\sl Nucleon Polarization in the Reaction $\pi^- p \to \pi^- \pi^+ n$}, Max Planck Institute for Physics and Astrophysics, Report MPI-PAE/Exp.El.75, 1978 (unpublished).

\bibitem{becker79a} H.~Becker {\sl et al.}, {\sl Measurement and Analysis of Reaction $\pi^- p \to \rho^0 n$ on Polarized Target}, Nucl.Phys. {\bf B150}, 301 (1979).

\bibitem{becker79b} H.~Becker {\sl et al.}, {\sl A Model Independent Partial-wave Analysis of the $\pi^+ \pi^-$ System Produced at Low Four-momentum Transfer in the Reaction $\pi^- p_{\uparrow} \to \pi^+ \pi^- n$ at 17.2 GeV/c}, Nucl.Phys. {\bf B151}, 46 (1979).

\bibitem{chabaud83} V.~Chabaud {\sl et al.}, {\sl Experimental Indications for a $2^{++}$ non-${\overline q}q$ Object}, Nucl.Phys. {\bf B223}, 1 (1983).

\bibitem{rybicki85} K.~Rybicki and I.~Sakrejda, {\sl Indication for a Broad $J^{PC}=2^{++}$ Meson at 840 MeV Produced in the Reaction $\pi^- p \to \pi^- \pi^+ n$ at High $|t|$}, Zeit.Phys. {\bf C28}, 65 (1985).

\bibitem{rybicki96} K.~Rybicki, {\sl Data Tables of Spin Moments Measured in $\pi^- p \to \pi^- \pi^+ n$ on Polarized Target at 17.2 GeV/c for Dipiom Masses 580 - 1600 MeV and Four-momentum Transfers $-t$=0.01 - 0.20 $(GeV/c)^2$}, private communication, 1996.

\bibitem{kaminski97} R.~Kami\'{n}ski, L.~Le\'{s}niak and K.~Rybicki, {\sl Separation of $S$-Wave Pseudoscalar and Pseudovector Amplitudes In $\pi^- p \to \pi^+\pi^- n$ On Polarized Target}, Zeit.Phys. {\bf C74}, 79 (1997).

\bibitem{kaminski02} R.~Kami\'{n}ski, L.~Le\'{s}niak and K.~Rybicki, {\sl A Joint Analysis of the $S$-wave in the $\pi^+\pi^-$ and $\pi^0\pi^0$ Data}, Eur.Phys.J.direct {\bf C4}, 1 (2002).

\bibitem{lesquen85} A.~de Lesquen, L.~van Rossum, M.~Svec {\sl et al.}, {\sl Measurement of the Reaction $\pi^+ n \to \pi^+ \pi^- p$ at 5.98 and 11.85 GeV/c Using a Transversely Polarized Deuteron Target}, Phys.Rev. {\bf D32}, 4355 (1985).

\bibitem{svec92a} M.~Svec, A.~de Lesquen and L.~ van Rossum, {\sl Amplitude Analysis of Reaction $\pi^+ n \to \pi^+ \pi^- p$ at 5.98 and 11.85 GeV/c}, Phys.Rev. {\bf D45}, 55 (1992).

\bibitem{alekseev99} I.G.~Alekseev {\sl et al.}(ITEP Collaboration), {\sl Study of the Reaction $\pi^- p \to \pi^- \pi^+ n$ on the Polarized Proton Target at 1.78 GeV/c: Experiment and Amplitude Analysis}, Nucl.Phys. {\bf B541}, 3 (1999).

\bibitem{donohue79} J.T.~Donohue and Y.~Leroyer, {\sl Is There a Narrow $\sigma$ under $\rho^0$ ?}, Nucl.Phys. {\bf B158}, 123 (1979).

\bibitem{svec84} M.~Svec, in {\sl SPIN 84}, {\sl Observation of a $0^{++}(750)$ Gluonium Candidate in Measurements of $\pi^+ n \to \pi^+ \pi^- p$ on Polarized Target at 5.98 and 11.85 GeV/c}, J.Phys. (Paris) Colloq. {\bf46}, C2 - 281 (1985).

\bibitem{svec92c} M.~Svec, A.~de Lesquen and L.~van Rossum, {\sl Evidence for a Scalar State I=0 $0^{++}(750)$ from Measurements of $\pi N \to \pi^+ \pi^- N$ on a Polarized Target at 5.98, 11.85 and 17.2 GeV/c}, Phys.Rev. {\bf D46}, 949 (1992).

\bibitem{svec96} M.~Svec, {\sl Study of $\sigma(750)$ and $\rho^0(770)$ Production in Measurements of $\pi N \to \pi^+ \pi^- N$ on a Polarized Target at 5.98, 11.85 and 17.2 GeV/c}, Phys.Rev. {\bf D53}, 2343 (1996).

\bibitem{svec97a} M.~Svec, {\sl Mass and Width of the $\sigma(750)$ Scalar Meson from Measurements of $\pi N \to \pi^- \pi^+ N$ on Polarized Target}, Phys.Rev. {\bf D55}, 5727 (1997).

\bibitem{svec02a}  M. ~Svec, {\sl Evidence for a Narrow $\sigma(770)$ Resonance and its Suppression in $\pi \pi$ Scattering from  Measurements of $\pi^- p \to \pi^- \pi^+ n$ on Polarized Target at 17.2 GeV/c}, hep-ph/0210249, 2002.

\bibitem{gunter01} J.~Gunter {\sl et al.} (BNL E852 Collaboration),
{\sl Partial Wave Analysis of the $\pi^0 \pi^0$ System Produced in $\pi^- p$ Charge Exchange Collisions}, Phys.Rev. {\bf D64}, 072003 (2001).

\bibitem{svec12d} M.~Svec, {\sl Consistency Tests of $\rho^0(770)-f_0(980)$ Mixing in $\pi^- p \to \pi^- \pi^+ n$}, arXiv:1411.2792 [hep-ph], 2014.

\bibitem{grayer74} G.~Grayer {\sl et al.}, {\sl High Statistics Study of the Reaction $\pi^- p \to \pi^- \pi^+ n$: Apparatus, Method of Analysis, and General Features of Results at 17.2 GeV/c}, Nucl.Phys. {\bf B75}, 189 (1974).

\bibitem{eadie83} W.T.~Eadie, {\sl Statistical Methods in Experimental Physics}, Elsevier Science, 1983.

\bibitem{bronshtein54} I.I.~Bronshtein, K.A.~Semendyayev, G.~Musiol and H.~M\"{u}hlig, {\sl Handbook of Mathematics}, Springer, 2007.

\bibitem{chin77} S.A.~Chin, {\sl A Relativistic Many-Body Theory of High Density Matter}, Annals of Physics (NY) {\bf 108}, 301 (1977).

\bibitem{gale00} O.~Teodorescu, A.K.~Muzunder and Ch.~Gale, {\sl Matter Induced $\rho-\delta$ Mixing: A Source of Dileptons}, Phys.Rev. {\bf C61}, 051901 (2000). 

\bibitem{gale01} O.~Teodorescu, A.K.~Muzunder and Ch.~Gale, {\sl Effects of Meson Mixing on Dilepton Spectra}, Phys.Rev. {\bf C63}, 034903  (2001).

\bibitem{gale02a} O.~Teodorescu, A.K.~Muzunder and Ch.~Gale, {\sl Aspects of Meson Properties in Dense Nuclear Matter}, Phys.Rev. {\bf C66}, 015209 (2002).

\bibitem{gale02b} O.~Teodorescu, A.K.~Muzunder and Ch.~Gale, {\sl Meson Mixing and Dilepton Production in Heavy Ion Collisions}, AIP Conf.Proc. {\bf 549}, 369 (2002).

\bibitem{svec90} M.~Svec, A de Lesquen and L. van Rossum, {\sl Spin Dependence of Matter Creation in Hadron Collisions}, Phys.Rev. {\bf 42}, 934 (1990).

\bibitem{svec12b} M.~Svec, {\sl Study of $\pi N \to \pi \pi N$ Processes on Polarized Targets I: Quantum Environment and Its Dephasing Interaction with Particle Scattering}, arXiv:1304.5120 [hep-ph], 2013.

\bibitem{svec13b} M.~Svec, {\sl Study of $\pi N \to \pi \pi N$ Processes on Polarized Targets II: The Prediction of $\rho^0(770)-f_0(980)$ Spin Mixing},
arXiv:1304.5400 [hep-ph], 2013.


\end{thebibliography}
\end{document}